\documentclass[amstex,12pt]{article}
\usepackage{amssymb}
\usepackage{amsmath}
\usepackage{amsthm}
\usepackage{braket}
\usepackage{mathtools}

\usepackage[normalem]{ulem}
\usepackage{hhline}
\usepackage{multirow,bigdelim}
\usepackage{graphicx}
\usepackage{enumitem}

\numberwithin{equation}{section}

\newcommand{\be}{\begin{equation}}
\newcommand{\ee}{\end{equation}}
\newcommand{\bea}{\begin{eqnarray}}
\newcommand{\eea}{\end{eqnarray}}
\newcommand{\non}{\nonumber}

\newcommand{\tr}{\mathop{\rm tr}\nolimits}
\newcommand{\diag}{\mathop{\rm diag}\nolimits}

\newcommand{\ff}{\mathfrak{f}}
\newcommand{\gcal}{\mathfrak{g}}
\newcommand{\ccal}{\mathfrak{c}}

\newcommand{\rcal}{\tt{r}}

\usepackage[usenames,dvipsnames]{color}

\newtheorem*{prop*}{Proposition}

\newtheorem*{corollary*}{Corollary}

\newcommand\blfootnote[1]{%
  \begingroup
  \renewcommand\thefootnote{}\footnote{#1}%
  \addtocounter{footnote}{-1}%
  \endgroup
}

\begin{document}

\begin{titlepage}
	\strut\hfill UMTG--298
	\vspace{.5in}
	\begin{center}
		
		\LARGE The spectrum of 
		quantum-group-invariant transfer matrices\\
		\vspace{1in}
		\large 
		Rafael I. Nepomechie \footnote{
			Physics Department,
			P.O. Box 248046, University of Miami, Coral Gables, FL 33124 USA}
		and Ana L. Retore ${}^{1,}$\footnote{
			Instituto de F\'{i}sica Te\'{o}rica-UNESP, Rua 
			Dr. Bento Teobaldo Ferraz 271, Bloco II 01140-070, S\~{a}o Paulo, Brazil}\\[0.8in]
	\end{center}
	
	\vspace{.5in}
	
\begin{abstract} 
Integrable open quantum spin-chain transfer matrices constructed from
trigonometric R-matrices associated to affine Lie algebras $\hat g$,
and from certain K-matrices (reflection matrices) depending on a
discrete parameter $p$, were recently considered in arXiv:1802.04864
and arXiv:1805.10144.  It was shown there that these transfer matrices
have quantum group symmetry corresponding to removing the $p^{th}$
node from the $\hat g$ Dynkin diagram.  Here we determine the spectrum
of these transfer matrices by using analytical Bethe ansatz, and we
determine the dependence of the corresponding Bethe equations on $p$.
We propose formulas for the Dynkin labels of the Bethe states in
terms of the numbers of Bethe roots of each type.We 
also briefly study how duality transformations are implemented on the Bethe ansatz 
solutions.
\end{abstract}
	
\blfootnote{e-mail addresses: {\tt nepomechie@miami.edu, ana.retore@unesp.br}}

\end{titlepage}

\setcounter{footnote}{0}

\section{Introduction}\label{sec:intro}

Several infinite families of integrable open quantum
spin chains with quantum group (QG) symmetry have recently been
identified \cite{Nepomechie:2018dsn, Nepomechie:2018wzp}.  The
transfer matrices for these models are constructed
\cite{Sklyanin:1988yz} from trigonometric R-matrices, which are
associated to non-exceptional affine Lie algebras $\hat g$
\cite{Bazhanov:1984gu, Bazhanov:1986mu, Jimbo:1985ua, Kuniba:1991yd}, and from certain
K-matrices (also known as reflection matrices, or boundary
S-matrices) depending on a discrete parameter $p$
\cite{Mezincescu:1990ui, Batchelor:1996np, Martins:2000xie,
LimaSantos:2002ui, LimaSantos:2003hx, Malara:2004bi,
Nepomechie:2018wzp}.  These transfer matrices have QG symmetry
corresponding to removing the $p^{th}$ node from the $\hat g$ Dynkin
diagram, as summarized in Tables \ref{table: QG symmetry general,
p=0,...,n} and \ref{table: QG symmetries special cases}.

The main aim of this paper is to determine the spectrum of these
transfer matrices.  This work can be regarded as a generalization of
the well-known work by Reshetikhin \cite{Reshetikhin:1987}, who
solved the corresponding problem for closed spin chains with periodic boundary
conditions, which however do {\em not} have QG symmetry.  Following
\cite{Reshetikhin:1987}, we use analytical Bethe ansatz to determine
the eigenvalues of the transfer matrices and the corresponding Bethe
equations.  We expect that the Bethe states are highest/lowest weights
of the quantum groups, which leads to formulas for the Dynkin labels
of the Bethe states in terms of the numbers of Bethe roots of each
type.  Analogous formulas have been known for integrable closed spin
chains constructed from {\em rational} R-matrices, which have
classical (Lie group) symmetries, see e.g. \cite{Reshetikhin:1986vd}.
From knowledge of the Dynkin labels of an irreducible representation,
one can determine its dimension, which in turn helps determine the
degeneracy of the corresponding transfer-matrix eigenvalue.

The K-matrices that we consider here, which do not depend on continuous 
boundary parameters, presumably correspond to conformal boundary 
conditions. Hence, these models may have interesting applications to 
boundary critical phenomena, see e.g. \cite{Batchelor:1995gx, 
Vernier:2014uja}.

The outline of this paper is as follows.  The construction of the
transfer matrix and key results from \cite{Nepomechie:2018dsn,
Nepomechie:2018wzp} are briefly reviewed in Sec. \ref{sec:review}. 
Expressions for the eigenvalues of the transfer matrix and 
corresponding Bethe equations are obtained in Sec. 
\ref{sec:Bethe Ansatz}.  Formulas for the Dynkin labels of the Bethe 
states (in terms of the numbers of Bethe roots of each type) are 
obtained and illustrated with some examples in Sec. \ref{sec:DynkinLabels}.
 We briefly study how duality transformations are implemented on the Bethe ansatz 
solutions in Sec.  \ref{sec:dualityBA}.
Some interesting open problems are listed in Sec. 
\ref{sec:discussion}. A connection between ``bonus'' symmetry and 
singular solutions of the Bethe equations is noted in Appendix 
\ref{sec:bonus}, and some additional cases are considered in Appendix 
\ref{sec:extra}.

\section{Review of previous results}\label{sec:review}

We briefly review here the construction of the transfer matrix 
(whose main ingredients are the R-matrix and the K-matrices)
and its symmetries.

\subsection{R-matrix}\label{sec:Rmat}

The R-matrix $R(u)$, which encodes the bulk interactions, is a
solution of the Yang-Baxter equation
\be
R_{12}(u - v)\,  R_{13}(u)\, R_{23}(v) = R_{23}(v)\, R_{13}(u)\, R_{12}(u - v)
\,,  \label{YBE}
\ee
\noindent
where $R_{12}=R\otimes \mathbb{I}$, $R_{23}=\mathbb{I}\otimes R$ and 
$R_{13}= {\cal P}_{12}\, R_{23}\, {\cal P}_{12}$; moreover, $\mathbb{I}$ 
is the identity matrix and ${\cal P}$ is 
the permutation matrix. We consider here the trigonometric R-matrices 
given by Jimbo \cite{Jimbo:1985ua} 
(except for $A_{2n-1}^{(2)}$, in which case we consider instead  
Kuniba's R-matrix \cite{Kuniba:1991yd}),
corresponding to the following non-exceptional affine Lie 
algebras\footnote{We do not consider here the case $A_n^{(1)}$, since
this R-matrix does not have crossing symmetry for $n>1$. This case has 
been studied in a similar context in \cite{deVega:1993ae, devega:1994hf, Doikou:1998ek}.}
\begin{equation}
\hat{g}=\left\{ A_{2n-1}^{(2)}\,, A_{2n}^{(2)}\,, B_n^{(1)}\,, 
C_n^{(1)}\,, D_n^{(1)}\,, D_{n+1}^{(2)}\right\} \,.
\label{hatg}
\end{equation}
We use the specific expressions for the R-matrices given in Appendix A of 
\cite{Nepomechie:2018dsn} and (for $D_{n+1}^{(2)}$) Appendix A of 
\cite{Nepomechie:2017hgw}, where the anisotropy parameter is denoted 
by $\eta$. We emphasize that, as in 
\cite{Nepomechie:2018dsn, Nepomechie:2018wzp}, we consider here 
exclusively generic values of $\eta$.
Various useful parameters related to these 
R-matrices are collected in Table \ref{table:definitions}. In 
particular, $d$ is the dimension of the vector space at each site of 
the spin chain; hence, the R-matrix is a $d^{2} \times d^{2}$ matrix.
Also, $\delta=0$ ($\delta=2$) for the untwisted 
(twisted) cases, respectively.

\begin{table}[h]
	\centering
	{\renewcommand{\arraystretch}{1.4}
		\begin{tabular}{|c|c|c|c|c|c|c|}
			\hline
			$ \hat{g} $ & $ A_{2n-1}^{(2)} $ & $ A_{2n}^{(2)} $ & $ B_{n}^{(1)} $ & $ C_{n}^{(1)} $ & $ D_{n}^{(1)} $ & $ D_{n+1}^{(2)} $\\
			\hline
			$ d $ & $ 2n $ & $ 2n+1 $& $ 2n+1 $& $ 2n $& $ 2n $& $ 2n+2 $\\
			\hline
			$ \kappa $ & $ 2n $ & $ 2n+1 $& $ 2n-1 $& $ 2n+2 $& $ 2n-2 $& $ 2n $\\
			\hline
			$ \rho $ & $ -2\kappa\eta-i\pi $ & $-2\kappa\eta-i\pi $& $ -2\kappa\eta $& $ -2\kappa\eta $& $ -2\kappa\eta $& $ -\kappa\eta $\\		
			\hline
			$ \omega $ & $ \kappa+2 $ & $  \kappa-2 $ & $  \kappa+2 $& $  \kappa-2 $& $ \kappa+2 $& $ \kappa $\\
			\hline
			$ \bar{\omega} $ & $ \kappa-2 $ & $  \kappa+2 $ & $  \kappa-2 $& $  \kappa+2 $& $ \kappa-2 $& $ \kappa $\\
			\hline
			$ \delta $ & $ 2 $ & $  2 $ & $ 0 $& $  0 $& $ 0 $& $ 2 $\\
			\hline
			$ \xi $ & $ 1 $ & $  0 $ & $ 0 $& $  1 $& $ 0 
			$& $ 0 $\\
			\hline
			$ \xi' $ & $ 0 $ & $  0 $ & $ 0 $& $  0 $& $ 1 
			$& $ 0 $\\
			\hline
	\end{tabular}}
	\caption{Parameters related to the R-matrices.}\label{table:definitions}
\end{table}

\subsection{K-matrices}\label{sec:Kmat}

The right ($K^{R}(u)$) and left ($K^{L}(u)$) K-matrices, which encode the boundary conditions on 
the right and left ends of the spin chain, respectively, are solutions 
of the boundary Yang-Baxter equations \cite{Cherednik:1985vs, Ghoshal:1993tm,
Sklyanin:1988yz, Mezincescu:1990uf}
\be
R_{12}(u - v)\, K^{R}_1(u)\, R_{21} (u + v)\, K^{R}_2(v)
= K^{R}_2(v)\, R_{12}(u + v)\, K^{R}_1(u)\, R_{21}(u - v)\,,
\label{BYBEm}
\ee
\noindent 
and
\begin{align}
\lefteqn{R_{12}(-u + v)\, K_1^{L\, t_1}(u)\, M^{-1}_1\, R_{21} (-u -v 
	-2\rho)\, M_1\, K_2^{L\, t_2}(v)} \non \\
&  \quad = K^{L\, t_2}_2(v)\, M_1\, R_{12}(-u - v- 2\rho)\, M^{-1}_1\,
K^{L\, t_1}_1(u)\, R_{21}(-u +v)  \,,
\label{BYBEp}
\end{align} 
respectively. The crossing parameter $\rho$ is given in Table 
\ref{table:definitions}, and the matrix $M$ can be found in \cite{Nepomechie:2018dsn} and (for $D_{n+1}^{(2)}$) 
\cite{Nepomechie:2017hgw}. 

For all the cases except $D_{n+1}^{(2)}$, we take for the right 
K-matrices the following $d \times d$ diagonal matrices \cite{Batchelor:1996np, 
LimaSantos:2002ui, LimaSantos:2003hx, Malara:2004bi}
\be
K^{R}(u,p) = \diag \big( \underbrace{e^{-u}\,, \ldots\,, 
	e^{-u}}_{p}\,, \underbrace{\frac{\gamma e^{u} + 1}{\gamma  + 
		e^{u}}\,, \ldots\,, \frac{\gamma e^{u} + 1}{\gamma  + 
		e^{u}}}_{d-2p}\,, \underbrace{e^{u}\,, \ldots\,, 
	e^{u}}_{p}\big) \,,
\label{KRa}
\ee
where $p$ is a discrete parameter taking the values
\begin{align}
& p=0,...,n \quad \quad \hspace{2.7cm}\text{ for } \quad A_{2n}^{(2)}\,, C_{n}^{(1)},\nonumber\\
& p=0,...,n\,, \quad p\neq 1\,,  \quad \hspace{1.2cm}\text{ for }\quad A_{2n-1}^{(2)}\,, B_{n}^{(1)},\nonumber\\
& p=0,...,n\,, \quad p\neq 1\,, n-1\,,  \quad \text{ for }\quad 
D_{n}^{(1)} \,.
\label{KRa p's}
\end{align}
Moreover, $\gamma$ is defined by
\begin{align}
\gamma = \begin{cases}
\gamma_{0}\, e^{(4p-2)\eta + \frac{1}{2}\rho} \qquad \mbox{ for  } 
\quad A_{2n-1}^{(2)}\,, B_{n}^{(1)}\,, D_{n}^{(1)}  \\ 
\hspace{0.2cm}\\
\gamma_{0}\, e^{(4p+2)\eta + \frac{1}{2}\rho} \qquad \mbox{ for  }
\quad A_{2n}^{(2)}\,, C_{n}^{(1)}  
\end{cases}\,,
\label{gamma}
\end{align}
where $\gamma_{0}$ is another discrete parameter
\be
\gamma_{0} = \pm 1 \,.
\ee
It is convenient to define the corresponding parameter
\be
\varepsilon=\frac{1-\gamma_0}{2}\,,
\ee
which therefore can take the values $\varepsilon = 0, 1$.

Note that in (\ref{KRa p's}) (as well as in \cite{Nepomechie:2018dsn})
the following cases are excluded:
\begin{align}
& A_{2n-1}^{(2)} \quad \text{with } \quad p=1,\nonumber\\
& B_n^{(1)} \quad\hspace{0.3cm} \text{with }\quad  p=1,\nonumber\\
& D_n^{(1)} \quad\hspace{0.3cm} \text{with } \quad p=1\,, n-1 \,.
\label{special cases}
\end{align}
For these cases, to which we henceforth refer as 
``special'' cases, we take instead the following right K-matrices:
\begin{equation}
K^{R}(u,1)=\text{diag}(e^{-2u},\underbrace{1,...,1}_{d-2},e^{2u})
\label{Kspecialp1}
\end{equation}
\noindent
for $A_{2n-1}^{(2)}\,, B_n^{(1)}\,, D_n^{(1)}$ with $p=1$; and
\begin{equation}
K^R(u,n-1)=\text{diag}(\,\underbrace{e^{-u},...,e^{-u}}_{n-1}\,,e^{u}\,,e^{-u},\,\underbrace{e^{u},...,e^{u}}_{n-1}\,) 
\label{Kspecialpn1}
\end{equation}
for $D_n^{(1)}$ with $p=n-1$. We choose these 
K-matrices because they lead to QG symmetry, as explained in Sec. 
\ref{subsec: Symmetries}.

For the case $D_{n+1}^{(2)}$, the right K-matrix is given by the $d
\times d$ block-diagonal matrix
\cite{Nepomechie:2018wzp}\footnote{The $D_{n+1}^{(2)}$ K-matrices
$K^R(u,n)$ (i.e. with $p=n$)
with $\varepsilon=0$ and $\varepsilon=1$ are proportional to the 
$D_{n+1}^{(2)}$ K-matrices
$K^-(u)$ in \cite{Nepomechie:2017hgw} for the cases I and II, 
respectively; explicitly, 
$K_{\cite{Nepomechie:2017hgw}\rm{I, 
II}}^{-}(u)=-2e^{2u+n\eta}\cosh\left(u-n\eta+\frac{i\pi}{2} 
\varepsilon\right)K^{R}(u,n)$.}
\begin{align}
	{\renewcommand{\arraystretch}{1.2}
K^{R}(u, p) = \left(\begin{array}{c|c|c c|c|c}
	k_{-}(u)\mathbb{I}_{p\times p} & & &  & &\\ \hline
	 & g(u)\mathbb{I}_{(n-p)\times (n-p)} & & & &  \\ \hline
	 & & k_1(u) & k_2(u) & &\\
	 & & k_2(u) & k_1(u) & &\\ \hline
	 & & & & g(u)\mathbb{I}_{(n-p)\times (n-p)} &\\ \hline
	 & & & & & k_{+}(u)\mathbb{I}_{p\times p}
\end{array}\right),
\label{KRb}}
\end{align}
\noindent
where 
\begin{align}
k_{\pm}(u) &=e^{\pm 2u}\,, \nonumber\\
g(u) &=\frac{\cosh\left(u-(n-2p)\eta+\frac{i 
\pi}{2}\varepsilon\right)}{\cosh\left(u+(n-2p)\eta-\frac{i 
\pi}{2}\varepsilon\right)}\,, \nonumber\\
k_1(u) &=\frac{\cosh(u)\cosh\left((n-2p)\eta+\frac{i 
\pi}{2}\varepsilon\right)}{\cosh\left(u+(n-2p)\eta+\frac{i 
\pi}{2}\varepsilon\right)}\,, \nonumber\\
k_2(u) &=-\frac{\sinh(u)\sinh\left((n-2p)\eta+\frac{i 
\pi}{2}\varepsilon\right)}{\cosh\left(u+(n-2p)\eta+\frac{i 
\pi}{2}\varepsilon\right)}\,.
\end{align}
and $\varepsilon$ is, again, a discrete parameter that can take the values $\varepsilon=0,1$.

Finally, for the left K-matrices, we take \cite{Sklyanin:1988yz, Mezincescu:1990uf}
\begin{equation}
K^{L}(u,p) = K^{R}(-u-\rho,p)\, M\,,
\label{KL}
\end{equation}
which is a solution of left boundary Yang-Baxter equation (\ref{BYBEp}), and corresponds to imposing the ``same'' boundary 
conditions on the two ends. 

\subsection{Transfer matrix}\label{sec:transfer}

The transfer matrix for an integrable open spin chain with $N$ sites 
is given by 
\cite{Sklyanin:1988yz}
\be
t(u,p) = \tr_a K^{L}_{a}(u,p)\, T_a(u)\,  K^{R}_{a}(u,p)\, \widehat{T}_a(u) \,, 
\label{transfer}
\ee
where the single-row monodromy matrices are defined by
\begin{align} 
T_a(u) &= R_{aN}(u)\ R_{a N-1}(u)\ \cdots R_{a1}(u) \,,  \non \\
\widehat{T}_a(u) &= R_{1a}(u)\ \cdots R_{N-1 a}(u)\ R_{Na}(u) \,,  
\label{monodromy}
\end{align}
and the trace in (\ref{transfer}) is over the ``auxiliary'' space, which 
is denoted by $a$. The transfer matrix satisfies the 
fundamental commutativity property
\be
\left[ t(u,p) \,, t(v,p) \right] = 0 \hbox{   for all   } u \,, v \,,
\label{commutativity} 
\ee
and contains the Hamiltonian ${\cal H}(p) \sim t'(0,p)$
as well as higher local conserved quantities. The transfer matrix is 
also crossing invariant
\be
t(u,p) = t(-u-\rho,p)\,,
\label{transfercrossing}
\ee
where the crossing parameter $\rho$ is given in Table \ref{table:definitions}. 

\subsection{Symmetries of the transfer matrix}\label{subsec: Symmetries}

It has been shown in \cite{Nepomechie:2018dsn, Nepomechie:2018wzp} 
that the transfer matrices (\ref{transfer}) constructed using the 
K-matrices \eqref{KRa} and \eqref{KRb} have the 
QG symmetries in Table \ref{table: QG symmetry general, p=0,...,n}. 
\begin{table}[h]
	\centering
	{\renewcommand{\arraystretch}{1.45}
	\begin{tabular}{|c|l|l|}
		\hline
		$\hat{g}$ & QG symmetry & Representation at each site\\
		\hline
		$A_{2n-1}^{(2)}$ & $U_q(C_{n-p})\otimes U_q(D_{p})$ $(p\neq 1)$ & $ (2(n-p),1)\oplus (1,2p)$\\
		\hline
		$A_{2n}^{(2)}$ & $U_q(B_{n-p})\otimes U_q(C_{p})$ & $(2(n-p)+1,1)\oplus (1,2p)$ \\
		\hline
		$B_{n}^{(1)}$ & $U_q(B_{n-p})\otimes U_q(D_{p})$  $(p\neq 1)$ &  $(2(n-p)+1,1)\oplus (1,2p)$\\
		\hline
		$C_{n}^{(1)}$ & $U_q(C_{n-p})\otimes U_q(C_{p})$ & $(2(n-p),1)\oplus (1,2p)$ \\
		\hline
		$D_{n}^{(1)}$ & $U_q(D_{n-p})\otimes U_q(D_{p})$ 
		$(n>1\,, p\neq 1\,, n-1)$ &  $(2(n-p),1)\oplus (1,2p)$ \\
		\hline
		$D_{n+1}^{(2)}$ & $U_q(B_{n-p})\otimes U_q(B_{p})$ &  
		$(2(n-p)+1,1)\oplus (1,2p+1)$ \\
		\hline
	\end{tabular}
	}
\caption{QG symmetries of the transfer matrix $t(u,p)$, where $p=0, 1, \ldots, 
n$.}\label{table: QG symmetry general, p=0,...,n}.
\end{table}
\noindent
For $0 < p < n$, the QG symmetries are given by a tensor product of 
two factors, to which we refer as the ``left'' and ``right'' factors.
For $p=0$, the ``right'' factors are absent; while for 
$p=n$, the ``left'' factors are absent.
That is,
\begin{align}
& \left[\Delta_{N}(H_{i}^{(l)}(p)) \,, t(u,p) \right] = 
\left[\Delta_{N}(E_{i}^{\pm (l)}(p)) \,, t(u,p) \right] = 0 
\,, \qquad i = 1\,, \ldots \,, n-p\,, \non \\
& \left[\Delta_{N}(H_{i}^{(r)}(p)) \,, t(u,p) \right] = 
\left[\Delta_{N}(E_{i}^{\pm (r)}(p)) \,, t(u,p) \right] = 0 
\,, \qquad i = 1\,, \ldots \,, p \,,
\label{QGsym}
\end{align}
where $H_{i}^{(l)}(p)\,, E_{i}^{\pm (l)}(p)$ are generators of the 
``left'' algebra $g^{(l)}$\,; $H_{i}^{(r)}(p)\,, 
E_{i}^{\pm (r)}(p)$ are generators of the ``right'' algebra 
$g^{(r)}$\,;
and $\Delta_{N}$ denotes the $N$-fold 
coproduct\footnote{The explicit form of $\Delta_N$ for $N=2$ can be 
found in \cite{Nepomechie:2018dsn,Nepomechie:2018wzp}}.

\begin{figure}[p]
	\includegraphics[width=16cm]{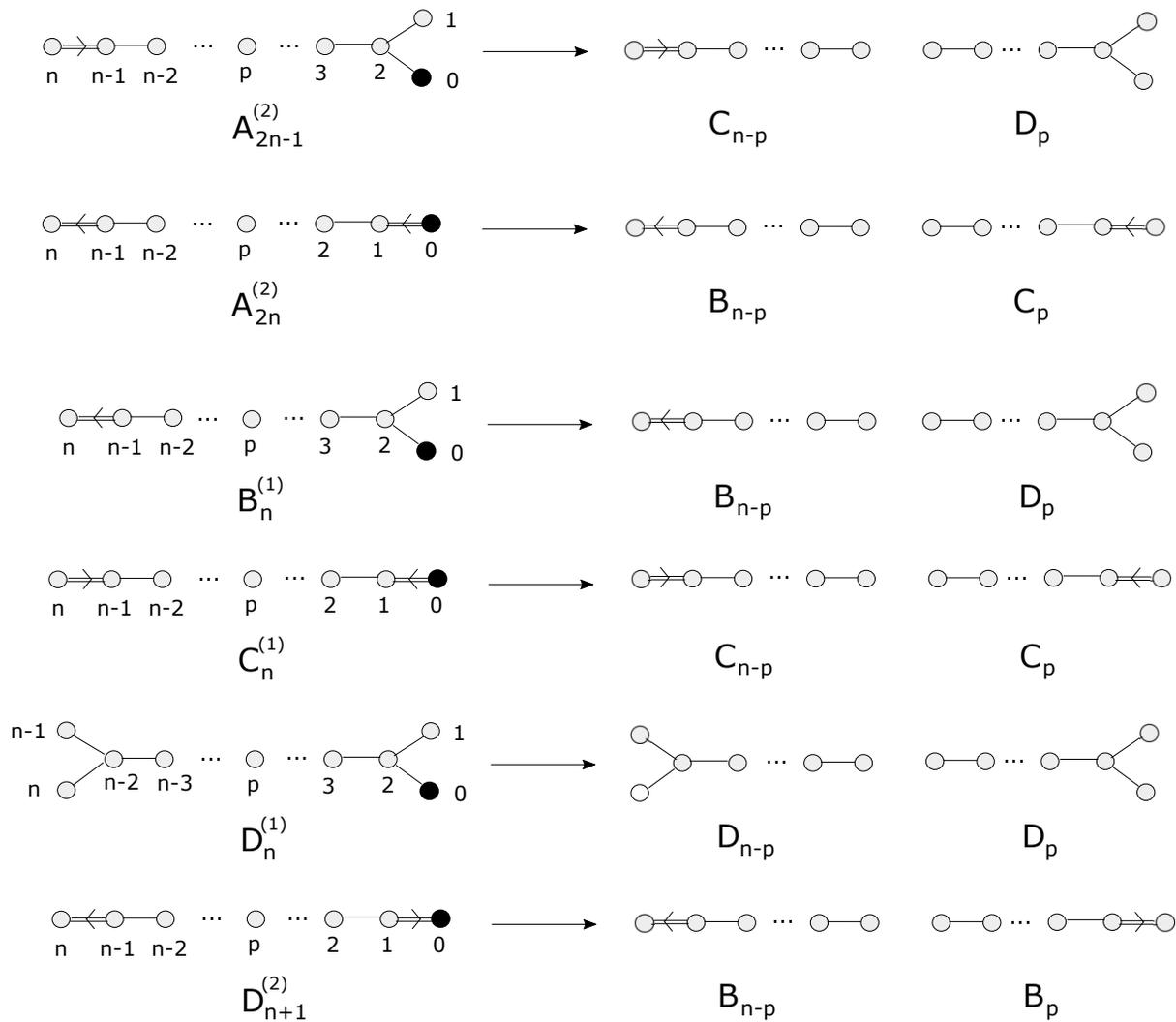}
	\caption{Subalgebras of affine Lie algebras corresponding 
	to removing the $p^{th}$ node from the extended Dynkin
	diagram.}\label{fig:Dynkin}
\end{figure}

It can be shown in a similar way that the transfer matrices
for the ``special'' cases (\ref{special cases}), which are constructed 
using the K-matrices (\ref{Kspecialp1})-(\ref{Kspecialpn1}),
have the QG symmetries in 
Table \ref{table: QG symmetries special cases}.\footnote{The 
symmetries for the ``special'' cases with $p=1$ are the same as for 
$p=0$, while the symmetry for $D_{n}^{(1)}$ with $p=n-1$ is the same 
as for $p=n$. (See Table \ref{table: QG symmetry general, p=0,...,n}.) 
These observations can be readily understood from the Dynkin 
diagrams, see Fig. \ref{fig:Dynkin}.\label{footnote: special cases}}

\begin{table}[h]
	\centering
	\renewcommand{\arraystretch}{1.4}
	\begin{tabular}{|c|c|c|}
		\hline
		$ \hat{g} $ & QG symmetry & Representation at each site\\
		\hline
		$ \hspace{-0.95cm}A_{2n-1}^{(2)}  (p=1) $  & $ U_q(C_{n}) $ & $ 2n $\\
		\hline
		$ \hspace{-1.1cm}B_{n}^{(1)}\,(p=1) $  & $ U_q(B_{n}) $ & $2n+1 $\\
		\hline
		$\hspace{1.2cm} D_{n}^{(1)}\, (n>1,p=1,n-1) $  & $ U_q(D_{n}) $ & $ 2n $\\
		\hline
	\end{tabular}
	\caption{QG symmetries of the transfer matrix $t(u,p)$ for 
	the ``special'' cases  (\ref{special cases}).}\label{table: QG symmetries special cases}
\end{table}

The QG symmetries displayed in Tables \ref{table: QG symmetry
general, p=0,...,n} and \ref{table: QG symmetries special cases}
correspond to removing the $p^{th}$ node from the $\hat{g}$ Dynkin 
diagram, as can be seen in Fig. \ref{fig:Dynkin}.

For the cases $C_{n}^{(1)}, D_{n}^{(1)}$ and $D_{n+1}^{(2)}$ (i.e., 
the last three rows of Table \ref{table: QG symmetry general, 
p=0,...,n}), the 
transfer matrices have a $p \leftrightarrow n-p$ duality symmetry
\be
{\cal U}\, t(u,p)\, {\cal U}^{-1} = f(u,p)\, t(u,n-p)  \,, 
\label{duality}
\ee 
see \cite{Nepomechie:2018dsn, Nepomechie:2018wzp} for explicit 
expressions for the quantum-space operator ${\cal U}$ and the scalar 
factor $f(u,p)$. In particular, for $p=\frac{n}{2}$ ($n$ even),
the transfer matrix is self-dual
\be
\left[ {\cal U}\,,  t(u, \tfrac{n}{2}) \right] = 0 \,,
\label{selfdual}
\ee
since $f(u,\frac{n}{2}) = 1$. For $p=\frac{n}{2}$ ($n$ even) and $\varepsilon=1$, there is an additional (``bonus'') 
symmetry, which leads to even higher degeneracies for the 
transfer-matrix eigenvalues \cite{Nepomechie:2018dsn, Nepomechie:2018wzp}. 

The cases $A_{2n-1}^{(2)}, B_{n}^{(1)}$ and $D_{n}^{(1)}$ (for 
which the ``right'' factor in Table \ref{table: QG symmetry general, 
p=0,...,n} is $U_{q}(D_{p})$) have a ``right'' $Z_{2}$ symmetry
\be
\left[{\cal Z}^{(r)}\,, t(u,p) \right] = 0 \,;
\label{Z2rightprop}
\ee 
and the case $D_{n}^{(1)}$ (for 
which the ``left'' factor in Table \ref{table: QG symmetry general, 
p=0,...,n} is $U_{q}(D_{n-p})$) also has a ``left'' $Z_{2}$ symmetry
\be
\left[{\cal Z}^{(l)}\,, t(u,p) \right] = 0 \,, 
\label{Z2leftprop}
\ee 
see \cite{Nepomechie:2018dsn} for explicit 
expressions for the quantum-space operators ${\cal Z}^{(r)}$ and 
${\cal Z}^{(l)}$.

\section{Analytical Bethe ansatz}\label{sec:Bethe Ansatz}

We now proceed to determine the spectrum of the transfer matrix 
(\ref{transfer}) for all the cases in Tables \ref{table: QG symmetry general, 
p=0,...,n} and \ref{table: QG symmetries special cases}. 
 The results hold for both values $\varepsilon=0, 1$ 
except for the case $D_{n+1}^{(2)}$, where we consider only $\varepsilon=0$.
The results 
for some of these cases have already been known: 
\begin{itemize}[noitemsep]
    \item For $p=0$: 
       \begin{itemize}[noitemsep]
          \item $A_{2n}^{(2)}$ \cite{Artz:1994qy, Li:2005pp, 
	  Li:2006mv, Ahmed:2017mqq} 
          \item $A_{2n-1}^{(2)}$ \cite{Artz:1995bm, Li:2006mv, Nepomechie:2017hgw}
	  \item $B_{n}^{(1)}$, $C_{n}^{(1)}$, 
	  $D_{n}^{(1)}$  \cite{Artz:1995bm, Li:2006mv}
	\end{itemize}  
    \item For $p=n$: 
       \begin{itemize}[noitemsep]
          \item $A_{2n}^{(2)}$ \cite{Li:2005pp, Ahmed:2017mqq}
          \item $A_{2n-1}^{(2)}$, $D_{n+1}^{(2)}$ \cite{Nepomechie:2017hgw}
      \end{itemize} 
    \item For $0 < p < n$:  
       \begin{itemize}[noitemsep]
	    \item $A_{2n}^{(2)}$ \cite{Li:2005pp}
       \end{itemize}
\end{itemize} 	  
The eigenvalues of the transfer matrix are determined in Secs. 
\ref{subsec:Eigenvalues}, \ref{subsec:yl}, and the corresponding Bethe 
equations are obtained in Sec. \ref{sec:Bethe equations}.

\subsection{Eigenvalues of the transfer matrix}\label{subsec:Eigenvalues}

The transfer matrix and Cartan generators can be diagonalized 
simultaneously 
\begin{align}
t(u,p)\, |\Lambda^{(m_{1}, \ldots\,, m_{n})}\rangle &= \Lambda^{(m_{1}, 
\ldots, m_{n})}(u,p)\,
|\Lambda^{(m_{1}, \ldots, m_{n})}\rangle \,, \non \\
\Delta_{N}(H_{i}^{(l)}(p))\, |\Lambda^{(m_{1}, \ldots, m_{n})}\rangle &= 
h^{(l)}_{i}\,
|\Lambda^{(m_{1}, \ldots, m_{n})}\rangle \,, \qquad i = 1, \ldots, 
n-p \,, \non \\
\Delta_{N}(H_{i}^{(r)}(p))\, |\Lambda^{(m_{1}, \ldots, m_{n})}\rangle &= 
h^{(r)}_{i}\,
|\Lambda^{(m_{1}, \ldots, m_{n})}\rangle \,, \qquad i = 1, \ldots, p \,, 
\label{eigenvalueproblem}
\end{align}
as follows from (\ref{commutativity}) and (\ref{QGsym}). We focus 
here on determining the transfer matrix eigenvalues $\Lambda^{(m_{1}, 
\ldots, m_{n})}(u,p)$; the 
eigenvalues of the Cartan generators $h^{(l)}_{i}, h^{(r)}_{i}$ are determined in Sec. 
\ref{subsec:Cartan}.

We take the analytical Bethe ansatz approach, whereby the eigenvalues 
of the transfer matrix are obtained by ``dressing'' the 
reference-state eigenvalue. The ``dressing''  is assumed to be 
``doubled'' with respect to the  corresponding closed chain. Hence, 
the main difficulty is to determine the 
reference-state eigenvalue. For the reference state corresponding to the cases in 
Table \ref{table: QG symmetry general, p=0,...,n}, we choose \footnote{ 
For the special cases in Table \ref{table: QG symmetries special 
cases}, we choose (see again footnote \ref{footnote: special cases}) the 
reference state $|0\,, 0\rangle$ for $A_{2n-1}^{(2)}$, 
$B_{n}^{(1)}$, $D_{n}^{(1)}$ with $p=1$; and the reference state 
$|0\,, n\rangle$ for $D_{n}^{(1)}$ with $p=n-1$.} 

\be
|0\,, p\rangle = v_{p}^{\otimes N}\,, \qquad 
v_{p} = \begin{cases}
e_{1} & \text{for}\quad p=0 \\
e_{p} &  \text{for}\quad p=1\,, \ldots\,, n
\end{cases}\,,
\label{reference}
\ee
where $e_{i}$ are $d$-dimensional elementary basis vectors $\left( e_{i} 
\right)_{j} = \delta_{i,j}$. Like the usual reference state 
$e_{1}^{\otimes N}$, the state (\ref{reference}) is an eigenstate of 
the transfer matrix with no Bethe roots ($m_{1} = \ldots = m_{n} = 0$)
\be
t(u,p)\, |0\,, p\rangle = \Lambda^{(0, \ldots, 0)}(u,p)\,|0\,, p\rangle \,.
\ee
However, in addition, this state is a highest weight of the ``left'' 
algebra 
\be
\Delta_{N}(E^{+ (l)}_{i}(p))\, |0\,, p\rangle =0 \,, 
\qquad i = 1, \ldots, n-p\,, 
\label{highestweightleftref}
\ee
and a lowest weight of the ``right'' algebra 
\be
\Delta_{N}(E^{- (r)}_{i}(p))\, |0\,, p\rangle =0 \,.
\qquad i = 1, \ldots, p\,.
\label{lowesttweightrightref}
\ee

In view of the crossing invariance (\ref{transfercrossing}) and the 
known results for $p=0$ \cite{Artz:1994qy, Artz:1995bm} and for 
$D_{n+1}^{(2)}$ \cite{Nepomechie:2017hgw},
we propose that the eigenvalues of the transfer matrix 
for general values of $p$ are given by the T-Q equation 

\begin{align}
\Lambda^{(m_1,...,m_n)}(u,p)=&\phi(u,p) \Bigg\{ A(u)\, z_0(u)\, 
y_0(u,p)\, 
c(u)^{2N}+\tilde{A}(u)\, \tilde{z}_0(u)\, \tilde{y}_0(u,p)\, \tilde{c}(u)^{2N}\nonumber\\
&\hspace{-0.3in} +\Big\{\sum_{l=1}^{n-1}\left[z_l(u)\, y_l(u,p)\, B_l(u)+\tilde{z}_l(u)\, 
\tilde{y}_l(u,p)\, \tilde{B}_l(u)\right]  + w_1(u)\, y_n(u,p)\, 
B_n(u)  \nonumber\\
&\hspace{-0.3in} + w_2\left[z_n(u)\, y_n(u,p)\, 
B_n(u)+\tilde{z}_n(u)\, \tilde{y}_n(u,p)\, 
\tilde{B}_n(u)\right]\Big\}\, b(u)^{2N} \Bigg\}\,.
\label{eigenvalue}
\end{align}
\noindent
The overall factor $\phi(u,p)$ is given by 
\be
\phi(u,p) = \begin{cases}
(-1)^{\xi} \left(\frac{\gamma e^{u} + 1}{\gamma + e^{u}}\right)\left(\frac{\gamma 
e^{-u-\rho} + 1}{\gamma + e^{-u-\rho}}\right) &\text{for}\quad  
A_{2n}^{(2)}\,, C_{n}^{(1)}\,,  A_{2n-1}^{(2)}\, (p\ne 1)\,, \\[-0.5\jot] 
&\quad
B_{n}^{(1)}\, (p\ne 1) 
\,, D_{n}^{(1)}\, (p\ne 1\,, n-1)\\
(-1)^{\xi}  &\text{for}\quad A_{2n-1}^{(2)}\, (p=1)\,, B_{n}^{(1)}\, 
(p=1) \,, \\[-0.5\jot] 
&\quad D_{n}^{(1)}\, (p= 1\,, n-1)\\
\frac{\cosh(u-(n-2p)\eta)\, 
\cosh(u-(n+2p)\eta)}{\cosh(u+(n-2p)\eta)\, \cosh(u-(3n-2p)\eta)}
&\text{for}\quad  D_{n+1}^{(2)}
\end{cases}
\label{phi}
\ee
where $\gamma$ is defined in (\ref{gamma}), and the parameters $\xi$ and  
$\rho$ are given in Table \ref{table:definitions}.
The tilde denotes crossing $\tilde{A}(u)=A(-u-\rho)$, etc.
The functions $A(u)$ and $B_l(u)$ for
$\hat{g}=A_{2n}^{(2)}$, $A_{2n-1}^{(2)}\, (\text{for } n>1)$, $B_{n}^{(1)}$, 
$C_{n}^{(1)}\, (\text{for } n>1)$, $D_{n}^{(1)}\, (\text{for } n>2)$ are defined as
\begin{align}
& A(u)=\frac{Q^{[1]}(u+2\eta)}{Q^{[1]}(u-2\eta)}\,, \nonumber \\
& B_l(u)=\frac{Q^{[l]}(u-2(l+2)\eta)}{Q^{[l]}(u-2l\eta)}\frac{Q^{[l+1]}(u-2(l-1)\eta)}{ Q^{[l+1]}(u-2(l+1)\eta)},\nonumber\\
& \hspace{1.5cm}l=1,...,n-3 \text{ for } D_n^{(1)}\nonumber\\
& \hspace{1.5cm}l=1,...,n-2 \text{ for } A_{2n-1}^{(2)}\,, C_n^{(1)}\nonumber\\
& \hspace{1.5cm}l=1,...,n-1 \text{ for } A_{2n}^{(2)}\,, B_n^{(1)} \,.
\label{A,B}
\end{align}
Moreover, for the values of $l$ not included above:
\begin{align}
& A_{2n-1}^{(2)}: 
& B_{n-1}(u) & =\frac{Q^{[n-1]}(u-2(n+1)\eta)}{Q^{[n-1]}(u-2(n-1)\eta)}
\frac{Q^{[n]}(u-2(n-2)\eta)}{Q^{[n]}(u-2n\eta)} \non\\
&&& \quad \times \frac{Q^{[n]}(u-2(n-2)\eta + i 
\pi)}{Q^{[n]}(u-2n\eta + i \pi)} \,,
\label{B A2n-1(2)} \\
& A_{2n}^{(2)}: 
& B_{n}(u) & =\frac{Q^{[n]}(u-2(n+2)\eta)}{Q^{[n]}(u-2n\eta)}\frac{Q^{[n]}(u-2(n-1)\eta+i \pi)}{Q^{[n]}(u-2(n+1)\eta+i\pi)} \,,
\label{B A2n(2)} 
\end{align}
\begin{align}
& B_{n}^{(1)}:
& B_{n}(u) & =\frac{Q^{[n]}(u-2(n-2)\eta)}{Q^{[n]}(u-2n\eta)}\frac{Q^{[n]}(u-2(n+1)\eta)}{Q^{[n]}(u-2(n-1)\eta)}\,,
\label{B Bn(1)} \\
& C_{n}^{(1)}:
& B_{n-1}(u) & =\frac{Q^{[n-1]}(u-2(n+1)\eta)}{Q^{[n-1]}(u-2(n-1)\eta)}\frac{Q^{[n]}(u-2(n-3)\eta)}{Q^{[n]}(u-2(n+1)\eta)}\,,
\label{B Cn(1)} 
\end{align}
\begin{align}
& D_{n}^{(1)}:
& B_{n-2}(u) & 
=\frac{Q^{[n-2]}(u-2n\eta)}{Q^{[n-2]}(u-2(n-2)\eta)}\frac{Q^{[n-1]}(u-2(n-3)\eta)}{Q^{[n-1]}(u-2(n-1)\eta)}\frac{Q^{[n]}(u-2(n-3)\eta)}{Q^{[n]}(u-2(n-1)\eta)}\,, \non\\
& &
B_{n-1}(u) & =\frac{Q^{[n-1]}(u-2(n-3)\eta)}{Q^{[n-1]}(u-2(n-1)\eta)}\frac{Q^{[n]}(u-2(n+1)\eta)}{Q^{[n]}(u-2(n-1)\eta)} \,.
\label{B D_n(1)}
\end{align}

For the values of $n$ not included above:
\begin{align}
A_1^{(2)}: \quad A(u) &=\frac{Q^{[1]}(u+2\eta)\, Q^{[1]}(u+2\eta + i 
\pi)}{Q^{[1]}(u-2\eta)\, Q^{[1]}(u-2\eta + i \pi)}\,,
\label{AA_1} \\
C_1^{(1)}: \quad A(u) &=\frac{Q^{[1]}(u+4\eta)}{Q^{[1]}(u-4\eta)}\,,
\label{AC_1} \\
D_2^{(1)}: \quad  
A(u) &=\frac{Q^{[1]}(u+2\eta)}{Q^{[1]}(u-2\eta)}\frac{Q^{[2]}(u+2\eta)}{Q^{[2]}(u-2\eta)}\,,\nonumber \\
B_1(u) &=\frac{Q^{[1]}(u-6\eta)}{Q^{[1]}(u-2\eta)}\frac{Q^{[2]}(u+2\eta)}{ Q^{[2]}(u-2\eta)}\,.
\end{align}
Finally, for $D_{n+1}^{(2)}$:
\begin{align}
& A(u)=\frac{Q^{[1]}(u+\eta)}{Q^{[1]}(u-\eta)}\frac{Q^{[1]}(u+\eta+i\pi)}{Q^{[1]}(u-\eta+i\pi)},\nonumber\\
& B_l(u)=\frac{Q^{[l]}(u-(l+2)\eta)}{Q^{[l]}(u-l\eta)}\frac{Q^{[l]}(u-(l+2)\eta+i\pi)}{Q^{[l]}(u-l\eta+i\pi)}\nonumber\\
& \hspace{1.2cm}\times \frac{Q^{[l+1]}(u-(l-1)\eta)}{Q^{[l+1]}(u-(l+1)\eta)}\frac{Q^{[l+1]}(u-(l-1)\eta+i\pi)}{Q^{[l+1]}(u-(l+1)\eta+i\pi)},\hspace{1cm} l=1,...,n-1,\nonumber\\
& B_n(u)=\frac{Q^{[n]}(u-(n+2)\eta)}{Q^{[n]}(u-n\eta)}\frac{Q^{[n]}(u-(n-2)\eta+i\pi)}{Q^{[n]}(u-n\eta+i\pi)} \,.
\label{B Dn(2)}
\end{align}
\noindent
In the above equations (\ref{A,B}) - (\ref{B Dn(2)}) for the functions $A(u)$ and $B_l(u)$, the functions 
$Q^{[l]}(u)$ are given by
\begin{equation}
Q^{[l]}(u)=\prod_{j=1}^{m_l}\sinh\left(\tfrac{1}{2}(u-u_j^{[l]})\right)\sinh\left(\tfrac{1}{2}(u+u_j^{[l]})\right)\,,
\qquad  Q^{[l]}(-u) = Q^{[l]}(u) \,,
\label{Q general}
\end{equation}
where the zeros $u_j^{[l]}$ (and their number $m_{l}$) are still to be determined. Note that 
these expressions for $A(u)$ and $B_l(u)$ are ``doubled'' with 
respect to those in \cite{Reshetikhin:1987} for the corresponding closed chains.

The functions $c(u)$ and $b(u)$ are given by
\begin{equation}
c(u)=\begin{cases}
2\sinh\left(\frac{u}{2}-2\eta\right)\cosh\left(\frac{u}{2}-\kappa\eta\right)&\quad \mbox{for}\quad A_{2n}^{(2)}\,, A_{2n-1}^{(2)}\,,\\
2\sinh\left(\frac{u}{2}-2\eta\right)\sinh\left(\frac{u}{2}-\kappa\eta\right)&\quad \mbox{for}\quad B_{n}^{(1)}\,, C_{n}^{(1)}\,, D_{n}^{(1)}\,,\\
4\sinh\left(u-2\eta\right)\sinh\left(u-\kappa\eta\right)&\quad 
\mbox{for}\quad D_{n+1}^{(2)}\,,
\end{cases}
\label{cfunctions}
\end{equation}
\noindent
and
\begin{equation}
b(u)=\begin{cases}
2\sinh\left(\frac{u}{2}\right)\cosh\left(\frac{u}{2}-\kappa\eta\right)&\quad \mbox{for}\quad A_{2n}^{(2)}\,, A_{2n-1}^{(2)}\,,\\
2\sinh\left(\frac{u}{2}\right)\sinh\left(\frac{u}{2}-\kappa\eta\right)&\quad \mbox{for}\quad B_{n}^{(1)}\,, C_{n}^{(1)}\,, D_{n}^{(1)}\,,\\
4\sinh\left(u\right)\sinh\left(u-\kappa\eta\right)&\quad 
\mbox{for}\quad D_{n+1}^{(2)}\,.
\end{cases}
\label{bfunctions}
\end{equation}

For all $\hat{g}$ except $D_{n+1}^{(2)}$, the functions $z_l(u)$ are given by 
\begin{equation}
z_l(u)=\frac{\sinh 
u\sinh(u-2\kappa\eta)\cosh\left(u-\omega\eta+(2-\delta)\frac{i\pi}{4}\right)}{\sinh(u-2l\eta)\sinh\left(u-2(l+1)\eta\right)
\cosh\left(u-\kappa\eta+(2-\delta)\frac{i\pi}{4}\right)}\,,
\label{z_l}
\end{equation}
where $\omega$ and $\delta$ are given in Table \ref{table:definitions}. 
For $D_{n+1}^{(2)}$ 
\begin{equation}
z_l(u)=\begin{cases}
\frac{\cosh(u-(n-1)\eta)\, \sinh(2u-4 n \eta)\, \sinh(u-(n+1)\eta) 
\sinh(2u)}{\sinh(u- n \eta)\, \cosh(u- n \eta)\, \sinh(2u-2l\eta)\, \sinh(2u-2(l+1)\eta)}
&  l=0,...,n-1\,, \\
z_{n-1}(u)\frac{\sinh\left(u-(n-1)\eta\right)}{\sinh\left(u-(n+1)\eta\right)} &  l=n \,.
\end{cases}
\end{equation}
\noindent

Finally, the quantities $w_1(u)$ and $w_{2}$ are defined as
\begin{align}
w_1(u) &=\begin{cases}
\frac{\sinh u\sinh(u-2\kappa\eta)}{\sinh(u-(\kappa+1)\eta) 
\sinh(u-(\kappa-1)\eta)} \quad & \mbox{for}\quad A_{2n}^{(2)}\,, B_n^{(1)}\,, \\
0\quad & \mbox{for} \quad A_{2n-1}^{(2)}\,, C_n^{(1)}\,, D_n^{(1)}\,, 
D_{n+1}^{(2)} \,,
\end{cases} \non \\
w_2 &=\begin{cases}
1 & \mbox{for}\quad D_{n+1}^{(2)} \\
0 & \mbox{otherwise}
\end{cases} \,.
\label{w1w2}
\end{align}
\noindent

In the expression (\ref{eigenvalue}) for the transfer-matrix
eigenvalue, only the functions $y_l(u,p)$ remain to be specified.  For
$y_l(u,p)=1$, the expression (\ref{eigenvalue}) reduces (apart from the 
overall factor) to the
transfer-matrix eigenvalue for the case $p=0$ for all the cases 
except $D_{n+1}^{(2)}$ \cite{Artz:1994qy, Artz:1995bm}.  
The functions $y_l(u,p)$ for general
values of $p$ are determined in the following section.

\subsection{Determining $y_l(u,p)$}\label{subsec:yl}

We now proceed to determine the functions $y_{0}(u,p)\,, \ldots\,, 
y_{n}(u,p)$ for general values
of $p$.  We emphasize that these are the only functions 
(besides the overall factor $\phi(u,p)$ (\ref{phi}), through the 
quantity $\gamma$ (\ref{gamma})) in the expression (\ref{eigenvalue}) for the transfer-matrix eigenvalue with explicit
dependence on $p$.

For the special cases in Table \ref{table: QG symmetries special 
cases}, the functions $y_l(u,p)$ are simply given by
\begin{equation}
y_l(u,p)=1 \,, \qquad l = 0\,, \ldots \,, n\,,
\label{y_l special cases}
\end{equation}
i.e., the same as for the case $p=0$.
We therefore focus our attention in the remainder of this section on the cases in Table \ref{table: QG symmetry general, p=0,...,n}.
 
We make the ansatz 
\begin{equation}
y_l(u,p) =\begin{cases}
F(u) \quad \mbox{for} \quad 0\le l \le p-1\\
G(u) \quad \mbox{for} \quad p\le l \le n
\end{cases}\,,
\label{yl}
\end{equation}
\noindent
and 
\begin{equation}
 \tilde{G}(u)\equiv G(-u-\rho)=G(u) \,,
 \label{G=Gtilde}
\end{equation} 
\noindent 
which guarantees that the only Bethe equation with an extra factor 
(in comparison with the case $p=0$) is the equation for the $p^{th}$ 
Bethe roots $\{ u^{[p]}_{j} \}$, as 
discussed further in Sec. \ref{sec:Bethe equations}. 

The explicit form of $F(u)$ and $G(u)$ are 
\begin{align}
& 
G(u)=\frac{\cosh\left(\frac{u}{2}-\frac{\omega\eta}{2}-(\delta-4\varepsilon)
\frac{i\pi}{8}\right)
\cosh\left(\frac{u}{2}-\frac{\bar{\omega}\eta}{2}-(\delta-4\varepsilon)\frac{i\pi}{8}\right)}
{\cosh\left(\frac{u}{2}-\frac{(\omega-4p)\eta}{2}-(\delta-4\varepsilon)\frac{i\pi}{8}\right)
\cosh\left(\frac{u}{2}-\frac{(\bar{\omega}+4p)\eta}{2}-(\delta-4\varepsilon)\frac{i\pi}{8}\right)}\,, \label{G1} \\
& 
F(u)=\frac{\cosh^2\left(\frac{u}{2}+\frac{(\omega-4p)\eta}{2}+(\delta-4\varepsilon)\frac{i\pi}{8}\right)}
{\cosh^2\left(\frac{u}{2}-\frac{\omega\eta}{2}-(\delta-4\varepsilon)\frac{i\pi}{8}\right)}\, G(u) \,,
\label{F1}
\end{align}
\noindent
for $\hat{g}=A_{2n}^{(2)}$, $A_{2n-1}^{(2)}$, $B_{n}^{(1)}$, 
$C_{n}^{(1)}$, $D_{n}^{(1)}$. Note that  $\omega\,, \bar\omega\,, 
\delta$ are given in Table \ref{table:definitions}. Moreover, 
\begin{align}
& G(u)=\frac{\cosh^{2}(u-n\eta)}{\cosh(u-(n-2p) \eta)\, 
\cosh(u-(n+2p) \eta)}\,, \label{G2} \\
& F(u)=\frac{\cosh^2\left(u+(n-2p)\eta\right)}{\cosh^2\left(u-n\eta\right)}\, G(u) \,,
\label{F2}
\end{align}
for $\hat{g}=D_{n+1}^{(2)}$. Note that $G(u)=1$ for 
$p=0$ in all cases.
The rest of this section is dedicated to explaining how the above 
expressions can be obtained, starting with $F(u)$.

According to (\ref{yl}), $y_0(u,p)$ is equal to $F(u)$ for any value of
$p$ except $p=0$.  We can use this fact to determine $F(u)$ by
arranging to kill all the terms in \eqref{eigenvalue} except the one with
$y_0(u,p)$, which can be accomplished by judiciously introducing 
inhomogeneities. Indeed, it is well known that arbitrary inhomogeneities $\{\theta_i\}$ can be introduced in the 
transfer matrix $t(u,p\,; \{\theta_i\})$ while maintaining the commutativity 
property
\begin{equation}
\left[t(u,p\,; \{\theta_i\})\,, t(v,p\,; \{\theta_i\})\right]=0 \,.
\label{commut inhomogeneities}
\end{equation}
\noindent
By appropriately choosing the inhomogeneities, all the terms in \eqref{eigenvalue}
except the first one can be made to vanish.  A similar procedure has 
been used in e.g. \cite{Nepomechie:2017hgw, Yung:1994tm}.

As an example, let us consider the case $A_{2n}^{(2)}$. The only 
effect on the eigenvalue \eqref{eigenvalue} of introducing inhomogeneities $\{\theta_i\}$ in the transfer 
matrix is to modify the expressions for $c(u)$, $\tilde{c}(u)$ and $b(u)$ 
(\ref{cfunctions}), (\ref{bfunctions}) as follows:
\begin{align}
&c(u)^{2N} 
=\left[2\sinh\left(\frac{u}{2}-2\eta\right)\cosh\left(\frac{u}{2}-\kappa\eta\right)\right]^{2N} \non\\
&\longmapsto \prod_{i=1}^{N}\left[2\sinh\left(\frac{u+\theta_i}{2}-2\eta\right)\cosh\left(\frac{u+\theta_i}{2}-\kappa\eta\right)\right]
\left[ 
2\sinh\left(\frac{u-\theta_i}{2}-2\eta\right)\cosh\left(\frac{u-\theta_i}{2}-\kappa\eta\right) \right]\,,\non
\end{align}
\begin{align}
&\tilde{c}(u)^{2N}
=\left[2\sinh\left(\frac{u}{2}\right)\cosh\left(\frac{u}{2}-\left(\kappa-2\right)\eta\right)\right]^{2N}\non\\
&\longmapsto \prod_{i=1}^{N}\left[2\sinh\left(\frac{u+\theta_i}{2}\right)\cosh\left(\frac{u+\theta_i}{2}-\left(\kappa-2\right)\eta\right)\right]
\left[ 
2\sinh\left(\frac{u-\theta_i}{2}\right)\cosh\left(\frac{u-\theta_i}{2}-\left(\kappa-2\right)\eta\right) \right]\,,\non
\end{align}
\begin{align}
&b(u)^{2N}
=\left[2\sinh\left(\frac{u}{2}\right)\cosh\left(\frac{u}{2}-\kappa\eta\right)\right]^{2N}\non \\
&\longmapsto 
\prod_{i=1}^{N}\left[2\sinh\left(\frac{u+\theta_i}{2}\right)\cosh\left(\frac{u+\theta_i}{2}-\kappa\,\eta\right)\right] 
\left[ 
2\sinh\left(\frac{u-\theta_i}{2}\right)\cosh\left(\frac{u-\theta_i}{2}-\kappa\,\eta\right) \right] \,. \non
\end{align}
By choosing $\theta_i=u$, the modified expressions for 
$\tilde{c}(u)$ and $b(u)$ (but not $c(u)$) evidently become zero; hence, the only 
term in \eqref{eigenvalue} that survives is the first term, which is 
proportional to $y_0(u,p)=F(u)$.
On the other hand, by acting with the transfer matrix 
$t(u,p\,;\{\theta_i=u\})$ with $N=1$ and $n=p=1$ on the
reference state (\ref{reference}),
we explicitly obtain the corresponding eigenvalue.  Comparing these two results, keeping in
mind that the reference state is the Bethe state with no Bethe roots
and therefore $A(u)=1$, we can solve for $F(u)$. By repeating this procedure 
for $n=2$ and $p=1, 2$, we infer the general result \eqref{F1},
which can then be easily checked in a similar way for higher values of $n, p, N$.

In order to determine $G(u)$, we return to the homogeneous case
$\theta_i=0$, so that all the functions $y_{0}(u,p), \ldots, y_{n}(u,p)$ again appear in
\eqref{eigenvalue}.  Using \eqref{eigenvalue}, the ans\"atze
\eqref{yl} and \eqref{G=Gtilde}, and the result \eqref{F1} for $F(u)$,
we obtain an expression for the reference-state eigenvalue
($A(u)=B_{l}(u)=1$) in terms of $G(u)$.  We also calculate this
eigenvalue explicitly by acting with $t(u,p)$ (with $N=1$) on the reference state
\eqref{reference}.  By comparing both expressions, we can solve for
$G(u)$.  We again use the results for small values of $n$ and $p$ to
infer the general result (\ref{G1}). Having obtained both $F(u)$ and $G(u)$ for general values of $n$ and
$p$, the reference-state eigenvalue can be easily checked for higher
values of $n, p, N$.

Using the same procedure for the other $\hat{g}$ in Table \ref{table: QG symmetry general, p=0,...,n}, we arrive at the 
results \eqref{G1} - \eqref{F2}.
As already noted, for the special cases in Table \ref{table: QG 
symmetries special cases}, we have $y_l(u,p)=1$ (\ref{y_l special 
cases}).

\subsection{Bethe equations}\label{sec:Bethe equations}

The expression (\ref{eigenvalue}) for the transfer-matrix eigenvalues
is in terms of the zeros $u_j^{[l]}$ of the functions $Q^{[l]}(u)$,
which are still to be determined.  In principle, these zeros can be
determined by solving corresponding Bethe equations, which we now
present.  We find that these Bethe equations are the same as for the
case $p=0$ \cite{Artz:1994qy, Artz:1995bm}, except
for the presence of an extra factor $\Phi_{l,p,n}(u)$ (\ref{phi1}), (\ref{phi2})  
that is different from 1 only if $l=p$. The only dependence on $p$ in the 
Bethe equations is in this factor.

\subsubsection{For $\hat{g}=A_{2n}^{(2)}\,, A_{2n-1}^{(2)}\,, 
B_{n}^{(1)}\,, C_{n}^{(1)}\,, D_{n}^{(1)}$}\label{subsubsec:BE1}

We determine the Bethe equations from the requirement that the expression
(\ref{eigenvalue}) for the transfer-matrix eigenvalues have vanishing
residues at the poles. In this way, we obtain the following Bethe equations
for all the cases in Tables \ref{table: QG
symmetry general, p=0,...,n} and \ref{table: QG symmetries special
cases} except for $D_{n+1}^{(2)}$:
\be
\left[\frac{\sinh\left(\frac{u_k^{[1]}}{2}+\eta\right)}
{\sinh\left(\frac{u_k^{[1]}}{2}-\eta\right)}\right]^{2N}\Phi_{1,p,n}(u_k^{[1]}) =
\frac{Q_k^{[1]}\left(u_k^{[1]}+4\eta\right)}{Q_k^{[1]}\left(u_k^{[1]}-4\eta\right)}\frac{Q^{[2]}\left(u_k^{[1]}-2\eta\right)}{Q^{[2]}
\left(u_k^{[1]}+2\eta\right)}\,, \quad k = 1, \ldots, m_{1}\,, \label{BEgen1}
\ee 
\begin{align}
&\Phi_{l,p,n}(u_k^{[l]})=\frac{Q^{[l-1]}\left(u_k^{[l]}-2\eta\right)}{Q^{[l-1]}
\left(u_k^{[l]}+2\eta\right)}\frac{Q_k^{[l]}\left(u_k^{[l]}+4\eta\right)}
{Q_k^{[l]}\left(u_k^{[l]}-4\eta\right)}\frac{Q^{[l+1]}\left(u_k^{[l]}-2\eta\right)}{Q^{[l+1]}\left(u_k^{[l]}+2\eta\right)}\,, 
\quad k = 1, \ldots, m_{l}\,,  \label{BEgen2} \\
& \qquad\qquad\qquad l=1,...,n-3 \quad \text{ for } \quad D_n^{(1)}\,(n>2) \non \\
& \qquad\qquad\qquad l=1,...,n-2 \quad \text{ for } \quad
C_n^{(1)}\, (n>1),\,A_{2n-1}^{(2)}\, (n>1) \non \\
& \qquad\qquad\qquad l=1,...,n-1 \quad \text{ for } \quad 
A_{2n}^{(2)}, B_n^{(1)} \,, \non
\end{align}
where $Q^{[l]}(u)$ is given by (\ref{Q general}), and $Q_k^{[l]}(u)$ is 
defined by
\begin{equation}
Q_k^{[l]}(u)=\prod_{j=1, j \ne 
k}^{m_l}\sinh\left(\tfrac{1}{2}(u-u_j^{[l]})\right)\sinh\left(\tfrac{1}{2}(u+u_j^{[l]})\right)\,.
\label{Q jk}
\end{equation}
Moreover, for the values of $l$ not included above:
\begin{align}
& A_{2n-1}^{(2)}:
   & \Phi_{n-1,p,n}(u_k^{[n-1]})& =\frac{Q^{[n-2]}\left(u_k^{[n-1]}-2\eta\right)}{Q^{[n-2]}\left(u_k^{[n-1]}+2\eta\right)}\frac{Q_k^{[n-1]}\left(u_k^{[n-1]}+4\eta\right)}{Q_k^{[n-1]}\left(u_k^{[n-1]}-4\eta\right)}\nonumber\\
&&&\quad \times 
\frac{Q^{[n]}\left(u_k^{[n-1]}-2\eta\right)}{Q^{[n]}\left(u_k^{[n-1]}+2\eta\right)}\frac{Q^{[n]}\left(u_k^{[n-1]}-2\eta+i\pi\right)}{Q^{[n]}\left(u_k^{[n-1]}+2\eta+i\pi\right)}\,,\label{BE extra A1}\\
&&\Phi_{n,p,n}(u_k^{[n]})&=\frac{Q^{[n-1]}\left(u_k^{[n]}-2\eta\right)}{Q^{[n-1]}\left(u_k^{[n]}+2\eta\right)}\frac{Q^{[n-1]}\left(u_k^{[n]}-2\eta+i\pi\right)}{Q^{[n-1]}\left(u_k^{[n]}+2\eta+i\pi\right)}\nonumber\\
&&&\quad \times 
\frac{Q_k^{[n]}\left(u_k^{[n]}+4\eta\right)}{Q_k^{[n]}\left(u_k^{[n]}-4\eta\right)}\frac{Q_k^{[n]}\left(u_k^{[n]}+4\eta+i\pi\right)}{Q_k^{[n]}\left(u_k^{[n]}-4\eta+i\pi\right)}\,, \label{BE extra A2}
\end{align}
\begin{equation}
A_{2n}^{(2)}: \qquad 
\Phi_{n,p,n}(u_k^{[n]})=\frac{Q^{[n-1]}\left(u_k^{[n]}-2\eta\right)}{Q^{[n-1]}\left(u_k^{[n]}+2\eta\right)}\frac{Q_k^{[n]}\left(u_k^{[n]}+4\eta\right)}{Q_k^{[n]}\left(u_k^{[n]}-4\eta\right)}
\frac{Q_k^{[n]}\left(u_k^{[n]}-2\eta+i\pi\right)}{Q_k^{[n]}\left(u_k^{[n]}+2\eta+i\pi\right)}\,, \label{BE extra AA}
\end{equation}
\begin{equation}
B_{n}^{(1)}: \qquad
\Phi_{n,p,n}(u_k^{[n]})=\frac{Q^{[n-1]}\left(u_k^{[n]}-2\eta\right)}{Q^{[n-1]}\left(u_k^{[n]}+2\eta\right)}
\frac{Q_k^{[n]}\left(u_k^{[n]}+2\eta\right)}{Q_k^{[n]}\left(u_k^{[n]}-2\eta\right)} \,,
\label{BE extra B}
\end{equation}
\begin{align}
    & C_{n}^{(1)}:
&\Phi_{n-1,p,n}(u_k^{[n-1]})&=\frac{Q^{[n-2]}\left(u_k^{[n-1]}-2\eta\right)}{Q^{[n-2]}\left(u_k^{[n-1]}+2\eta\right)}
\frac{Q_k^{[n-1]}\left(u_k^{[n-1]}+4\eta\right)}{Q_k^{[n-1]}\left(u_k^{[n-1]}-4\eta\right)}
\frac{Q^{[n]}\left(u_k^{[n-1]}-4\eta\right)}{Q^{[n]}\left(u_k^{[n-1]}+4\eta\right)}\,, \label{BE extra C1}\\
&&\Phi_{n,p,n}(u_k^{[n]})&=\frac{Q^{[n-1]}\left(u_k^{[n]}-4\eta\right)}{Q^{[n-1]}\left(u_k^{[n]}+4\eta\right)}
\frac{Q_k^{[n]}\left(u_k^{[n]}+8\eta\right)}{Q_k^{[n]}\left(u_k^{[n]}-8\eta\right)} \,,
\label{BE extra C2}
\end{align}
\begin{align}
    & D_{n}^{(1)}:
&\Phi_{n-2,p,n}(u_k^{[n-2]})&=\frac{Q^{[n-3]}\left(u_k^{[n-2]}-2\eta\right)}{Q^{[n-3]}\left(u_k^{[n-2]}+2\eta\right)}
\frac{Q_k^{[n-2]}\left(u_k^{[n-2]}+4\eta\right)}{Q_k^{[n-2]}\left(u_k^{[n-2]}-4\eta\right)} \nonumber\\
&&&\quad \times \frac{Q^{[n-1]}\left(u_k^{[n-2]}-2\eta\right)}{Q^{[n-1]}\left(u_k^{[n-2]}+2\eta\right)}
\frac{Q^{[n]}\left(u_k^{[n-2]}-2\eta\right)}{Q^{[n]}\left(u_k^{[n-2]}+2\eta\right)}\,, \label{BE extra D1}\\
&&\Phi_{n-1,p,n}(u_k^{[n-1]})&=\frac{Q^{[n-2]}\left(u_k^{[n-1]}-2\eta\right)}{Q^{[n-2]}\left(u_k^{[n-1]}+2\eta\right)}
\frac{Q_k^{[n-1]}\left(u_k^{[n-1]}+4\eta\right)}{Q_k^{[n-1]}\left(u_k^{[n-1]}-4\eta\right)}\,,
\label{BE extra D2}\\
&&\Phi_{n,p,n}(u_k^{[n]})&=\frac{Q^{[n-2]}\left(u_k^{[n]}-2\eta\right)}{Q^{[n-2]}\left(u_k^{[n]}+2\eta\right)}
\frac{Q_k^{[n]}\left(u_k^{[n]}+4\eta\right)}{Q_k^{[n]}\left(u_k^{[n]}-4\eta\right)}\,. \label{BE extra D3}
\end{align}

The Bethe equations for values of $n$ not included above:
\begin{align}
A_1^{(2)}: \quad    
\left[\frac{\sinh(\frac{u_k^{[1]}}{2}+2\eta)}{\sinh(\frac{u_k^{[1]}}{2}-2\eta)}\right]^{2N}\Phi_{1,p,1}(u_k^{[1]})
&=\frac{Q_k^{[1]}\left(u_k^{[1]}+4\eta\right)}{Q_k^{[1]}\left(u_k^{[1]}-4\eta\right)}
\frac{Q_k^{[1]}\left(u_k^{[1]}+4\eta + i 
\pi\right)}{Q_k^{[1]}\left(u_k^{[1]}-4\eta + i \pi\right)}\,,  \\
A_3^{(2)}: \quad  
\left[\frac{\sinh(\frac{u_k^{[1]}}{2}+\eta)}{\sinh(\frac{u_k^{[1]}}{2}-\eta)}\right]^{2N}\Phi_{1,p,2}(u_k^{[1]})
&=\frac{Q_k^{[1]}\left(u_k^{[1]}+4\eta\right)}{Q_k^{[1]}\left(u_k^{[1]}-4\eta\right)}
\frac{Q^{[2]}\left(u_k^{[1]}-2\eta\right)}{Q^{[2]}\left(u_k^{[1]}+2\eta\right)}
\frac{Q^{[2]}\left(u_k^{[1]}-2\eta + i 
\pi\right)}{Q^{[2]}\left(u_k^{[1]}+2\eta+ i \pi\right)}\,,   \non \\
\Phi_{2,p,2}(u_k^{[2]}) & 
= \frac{Q^{[1]}\left(u_k^{[2]}-2\eta\right)}{Q^{[1]}\left(u_k^{[2]}+2\eta\right)}
\frac{Q^{[1]}\left(u_k^{[2]}-2\eta+ i 
\pi\right)}{Q^{[1]}\left(u_k^{[2]}+2\eta+ i \pi\right)} \non \\
& \times 
\frac{Q_k^{[2]}\left(u_k^{[2]}+4\eta\right)}{Q_k^{[2]}\left(u_k^{[2]}-4\eta\right)}
\frac{Q_k^{[2]}\left(u_k^{[2]}+4\eta+ i 
\pi\right)}{Q_k^{[2]}\left(u_k^{[2]}-4\eta+ i \pi\right)}\,,
\end{align}
\be
A_2^{(2)}: \quad    
\left[\frac{\sinh(\frac{u_k^{[1]}}{2}+\eta)}{\sinh(\frac{u_k^{[1]}}{2}-\eta)}\right]^{2N}\Phi_{1,p,1}(u_k^{[1]})
=\frac{Q_k^{[1]}\left(u_k^{[1]}+4\eta\right)}{Q_k^{[1]}\left(u_k^{[1]}-4\eta\right)}
\frac{Q_k^{[1]}\left(u_k^{[1]}-2\eta + i 
\pi\right)}{Q_k^{[1]}\left(u_k^{[1]}+2\eta + i \pi\right)}\,, 
\ee 
\be
B_1^{(1)}: \quad    
\left[\frac{\sinh(\frac{u_k^{[1]}}{2}+\eta)}{\sinh(\frac{u_k^{[1]}}{2}-\eta)}\right]^{2N}\Phi_{1,p,1}(u_k^{[1]})
=\frac{Q_k^{[1]}\left(u_k^{[1]}+2\eta\right)}{Q_k^{[1]}\left(u_k^{[1]}-2\eta\right)}\,, 
\ee 
\begin{align}
C_1^{(1)}: \quad  
\left[\frac{\sinh(\frac{u_k^{[1]}}{2}+2\eta)}{\sinh(\frac{u_k^{[1]}}{2}-2\eta)}\right]^{2N}\Phi_{1,p,1}(u_k^{[1]})
& =\frac{Q_k^{[1]}\left(u_k^{[1]}+8\eta\right)}{Q_k^{[1]}\left(u_k^{[1]}-8\eta\right)}\,, \\
C_2^{(1)}: \quad 
\left[\frac{\sinh(\frac{u_k^{[1]}}{2}+\eta)}{\sinh(\frac{u_k^{[1]}}{2}-\eta)}\right]^{2N}\Phi_{1,p,2}(u_k^{[1]})
& 
=\frac{Q_k^{[1]}\left(u_k^{[1]}+4\eta\right)}{Q_k^{[1]}\left(u_k^{[1]}-4\eta\right)}
\frac{Q^{[2]}\left(u_k^{[1]}-4\eta\right)}{Q^{[2]}\left(u_k^{[1]}+4\eta\right)}\,,  \\
\Phi_{2,p,2}(u_k^{[2]}) & 
= \frac{Q^{[1]}\left(u_k^{[2]}-4\eta\right)}{Q^{[1]}\left(u_k^{[2]}+4\eta\right)}
\frac{Q_k^{[2]}\left(u_k^{[2]}+8\eta\right)}{Q_k^{[2]}\left(u_k^{[2]}-8\eta\right)} \,,
\end{align}
\begin{align}
D_2^{(1)} :\quad  
\left[\frac{\sinh(\frac{u_k^{[1]}}{2}+\eta)}{\sinh(\frac{u_k^{[1]}}{2}-\eta)}\right]^{2N}\Phi_{1,p,2}(u_k^{[1]}) &=
\frac{Q_k^{[1]}\left(u_k^{[1]}+4\eta\right)}{Q_k^{[1]}\left(u_k^{[1]}-4\eta\right)}\,,  \label{BED2a}\\
\left[\frac{\sinh(\frac{u_k^{[2]}}{2}+\eta)}{\sinh(\frac{u_k^{[2]}}{2}-\eta)}\right]^{2N}\Phi_{2,p,2}(u_k^{[2]}) &=
\frac{Q_k^{[2]}\left(u_k^{[2]}+4\eta\right)}{Q_k^{[2]}\left(u_k^{[2]}-4\eta\right)}\,,
\label{BED2b} \\
D_3^{(1)}: \quad  
\left[\frac{\sinh(\frac{u_k^{[1]}}{2}+\eta)}{\sinh(\frac{u_k^{[1]}}{2}-\eta)}\right]^{2N}\Phi_{1,p,3}(u_k^{[1]}) &=
\frac{Q_k^{[1]}\left(u_k^{[1]}+4\eta\right)}{Q_k^{[1]}\left(u_k^{[1]}-4\eta\right)}
\frac{Q^{[2]}\left(u_k^{[1]}-2\eta\right)}{Q^{[2]}\left(u_k^{[1]}+2\eta\right)} \non \\
&\quad \times \frac{Q^{[3]}\left(u_k^{[1]}-2\eta\right)}{Q^{[3]}\left(u_k^{[1]}+2\eta\right)} \,, \\
\Phi_{2,p,3}(u_k^{[2]}) &= 
\frac{Q^{[1]}\left(u_k^{[2]}-2\eta\right)}{Q^{[1]}\left(u_k^{[2]}+2\eta\right)}
\frac{Q_k^{[2]}\left(u_k^{[2]}+4\eta\right)}{Q_k^{[2]}\left(u_k^{[2]}-4\eta\right)}\,, \\
\Phi_{3,p,3}(u_k^{[3]}) &= 
\frac{Q_k^{[1]}\left(u_k^{[3]}-2\eta\right)}{Q_k^{[1]}\left(u_k^{[3]}+2\eta\right)}
\frac{Q_k^{[3]}\left(u_k^{[3]}+4\eta\right)}{Q_k^{[3]}\left(u_k^{[3]}-4\eta\right)}\,.
\end{align}
The $u_k^{[1]} \leftrightarrow u_k^{[2]}$ symmetry of the Bethe 
equations (\ref{BED2a}), (\ref{BED2b}) is a reflection of the 
$U_q\left(D_2\right)$ symmetry (see again Table \ref{table: QG
symmetry general, p=0,...,n}) and the fact $D_2=A_1\otimes A_1$.

The important factor $\Phi_{l,p,n}(u)$ in the Bethe 
equations for most of the cases in Table \ref{table: QG symmetry 
general, p=0,...,n} is given by \footnote{The exceptions are as 
follows:
\begin{align*}
    A^{(2)}_{2n-1}\,, p=n : \quad \Phi_{l,p,n}(u) &= \begin{cases}
    \frac{\tilde{y}_{n-1}(u + 2 n \eta,p)}{y_{n-1}(u + 2 n \eta,p)}  & \text{for } l=n \\
    1 & \text{for }  l\ne n
\end{cases} \,, \\
   C^{(1)}_{n}\,, p=n : \quad \Phi_{l,p,n}(u) &= \begin{cases}
    \frac{\tilde{y}_{n-1}(u + 2 (n+1) \eta,p)}{y_{n-1}(u + 2 (n+1) \eta,p)}  & 
    \text{for } l=n \\
    1 & \text{for }  l\ne n
\end{cases} \,, \\
   D^{(1)}_{n}\,, p=n : \quad \Phi_{l,p,n}(u) &= \begin{cases}
    \frac{y_{n-1}(u + 2 (n-1) \eta,p)}{y_{n-2}(u + 2 (n-1) \eta,p)}  & 
    \text{for } l=n \\
    1 & \text{for }  l\ne n
\end{cases} \,.
\end{align*}}
\be
\Phi_{l,p,n}(u)=\frac{y_{l}(u + 2 l \eta,p)}{y_{l-1}(u + 2 l \eta,p)} = 
\begin{cases}
    \frac{G(u + 2 p \eta)}{F(u + 2 p \eta)} & \text{for }  l=p \\
    1 & \text{for }  l\ne p
\end{cases} \,,
\ee 
where the second equality follows from  (\ref{yl}). Using the 
expressions for $G(u)$ (\ref{G1}) and  $F(u)$ (\ref{F1}), we conclude that 
$\Phi_{l,p,n}(u)$ is given by
\begin{equation}
\Phi_{l,p,n}(u)=\begin{cases}\left[\frac{\cosh\left(\frac{u}{2}-\delta_{l,p}\left(\frac{(\omega-2p)}{2}\eta+\frac{i\pi}{8}(\delta-4\varepsilon)\right)\right)}
{\cosh\left(\frac{u}{2}+\delta_{l,p}\left(\frac{(\omega-2p)}{2}\eta+\frac{i\pi}{8}(\delta-4\varepsilon)\right)\right)}\right]^2& 
\text{for }\quad B_n^{(1)}\,,  C_n^{(1)}\,, D_n^{(1)}\,, A_{2n}^{(2)} 
\\[-1.0\jot] 
& \text{and for}\quad  A_{2n-1}^{(2)} \text{ with } l<n\,,\\
\\[1pt]
\left[\frac{\sinh\left(u-\delta_{l,p}\left((\omega-2p)\eta+\frac{i\pi}{4}(\delta-4\varepsilon)\right)\right)}
{\sinh\left(u+\delta_{l,p}\left((\omega-2p)\eta+\frac{i\pi}{4}(\delta-4\varepsilon)\right)\right)}\right]^2 & \text{for } A_{2n-1}^{(2)} \text{ with } l=n\,.
\end{cases}
\label{phi1}
\end{equation}
\noindent
Note that $\Phi_{l,p,n}(u)$ is different from 1 only if $l=p$. That 
is,  the Bethe equations are the same as for the
case $p=0$ \cite{Artz:1994qy, Artz:1995bm}, 
except for an extra factor in the equation for the $p^{th}$ Bethe 
roots $\{ u^{[p]}_{j} \}$. 

The factor $\Phi_{l,p,n}(u)$ for all the special cases in Table \ref{table: QG symmetries special 
cases} is simply given by
\begin{equation}
\Phi_{l,p,n}(u)=1 \,,
\label{phi3}
\end{equation}
as follows from (\ref{y_l special cases}).

For $p=n$, the Bethe equations for 
$A_{2n}^{(2)}$ with $\varepsilon=1$ reduce to those found in 
\cite{Ahmed:2017mqq}; and (again for $p=n$) the Bethe equations for $A_{2n-1}^{(2)}$
with $\varepsilon=0$ reduce to those found in \cite{Nepomechie:2017hgw}.
We have numerically verified the completeness of all the above Bethe ansatz solutions for 
small values of $n$ and $N$ (for all $p=0, \ldots, n$ and 
$\varepsilon = 0, 1$), 
along the lines in \cite{Ahmed:2017mqq, 
Nepomechie:2017hgw}.

\subsubsection{For $\hat{g}=D_{n+1}^{(2)}$}\label{subsec:D2n}

We emphasize that, for $D_{n+1}^{(2)}$, we consider only the case 
$\varepsilon=0$. We obtain the following  Bethe equations:\\
\noindent
For $n=1$ with $p=0, 1$:
\begin{equation}
\left[\frac{\sinh(u_k^{[1]}+\eta)}{\sinh(u_k^{[1]}-\eta)}\right]^{2N}=\frac{Q_k^{[1]}\left(u_k^{[1]}+2\eta\right)}{Q_k^{[1]}\left(u_k^{[1]}-2\eta\right)}\,, 
\quad k = 1, \ldots, m_{1} \,.
\end{equation}
\noindent
For $n>1$ with $p=0, \ldots, n$:
\begin{align}
\left[\frac{\sinh(u_k^{[1]}+\eta)}{\sinh(u_k^{[1]}-\eta)}\right]^{2N}\Phi_{1,p,n}(u_k^{[1]})&=\frac{Q_k^{[1]}\left(u_k^{[1]}+2\eta\right)}{Q_k^{[1]}\left(u_k^{[1]}-2\eta\right)}\frac{Q_k^{[1]}\left(u_k^{[1]}+2\eta+i\pi\right)}{Q_k^{[1]}\left(u_k^{[1]}-2\eta+i\pi\right)}\nonumber\\
&\quad \times 
\frac{Q^{[2]}\left(u_k^{[1]}-\eta\right)}{Q^{[2]}\left(u_k^{[1]}+\eta\right)}\frac{Q^{[2]}\left(u_k^{[1]}-\eta+i\pi\right)}{Q^{[2]}\left(u_k^{[1]}+\eta+i\pi\right)}\,, \nonumber\\
&\quad k=1,\ldots, m_1\,,
\end{align}
\noindent
\begin{align}
\Phi_{l,p,n}(u_k^{[l]})& =\frac{Q^{[l-1]}\left(u_k^{[l]}-\eta\right)}{Q^{[l-1]}\left(u_k^{[l]}+\eta\right)}\frac{Q^{[l-1]}\left(u_k^{[l]}-\eta+i\pi\right)}{Q^{[l-1]}\left(u_k^{[l]}+\eta+i\pi\right)}\nonumber\\
&\quad \times \frac{Q_k^{[l]}\left(u_k^{[l]}+2\eta\right)}{Q_k^{[l]}\left(u_k^{[l]}-2\eta\right)}\frac{Q_k^{[l]}\left(u_k^{[l]}+2\eta+i\pi\right)}{Q_k^{[l]}\left(u_k^{[l]}-2\eta+i\pi\right)}\nonumber\\
&\quad \times 
\frac{Q^{[l+1]}\left(u_k^{[l]}-\eta\right)}{Q^{[l+1]}\left(u_k^{[l]}+\eta\right)}\frac{Q^{[l+1]}\left(u_k^{[l]}-\eta+i\pi\right)}{Q^{[l+1]}\left(u_k^{[l]}+\eta+i\pi\right)}\,, \nonumber\\
&\quad k=1, \ldots, m_l\,, \quad l=2,\ldots, n-1\,,
\end{align}
\noindent
\begin{align}
\Phi_{n,p,n}(u_k^{[n]})& 
=\frac{Q^{[n-1]}\left(u_k^{[n]}-\eta\right)}{Q^{[n-1]}\left(u_k^{[n]}+\eta\right)}\frac{Q^{[n-1]}\left(u_k^{[n]}-\eta+i\pi\right)}{Q^{[n-1]}
\left(u_k^{[n]}+\eta+i\pi\right)}\frac{Q_k^{[n]}\left(u_k^{[n]}+2\eta\right)}{Q_k^{[n]}\left(u_k^{[n]}-2\eta\right)}\,, \nonumber\\
&\quad k=1, \ldots, m_n \,.
\end{align}
The factor $\Phi_{l,p,n}(u)$ in the above Bethe equations is 
given by
\be
\Phi_{l,p,n}(u)=\frac{y_{l}(u + l \eta,p)}{y_{l-1}(u + l \eta,p)} = 
\begin{cases}
    \frac{G(u + p \eta)}{F(u + p \eta)} & \text{for }  l=p \\
    1 & \text{for }  l\ne p
\end{cases} \,.
\ee 
Using the results for $G(u)$ (\ref{G2}) and  $F(u)$ (\ref{F2}), 
we obtain
\begin{equation}
\Phi_{l,p,n}(u)=\left[\frac{\cosh\left(u-\delta_{l,p}(n-p)\eta\right)}
{\cosh\left(u+\delta_{l,p}(n-p)\eta\right)}\right]^2 \,.
\label{phi2}
\end{equation}
As for (\ref{phi1}), this factor $\Phi_{l,p,n}(u)$ is 
different from 1 only if $l=p$.

For $p=n$, these Bethe equations reduce to the one found in 
\cite{Nepomechie:2017hgw}.
We have numerically verified the completeness of the above Bethe 
ansatz solutions for small values of $n$ and $N$ (for all $p=0, 
\ldots, n$) along the lines in \cite{Nepomechie:2017hgw}.


\subsubsection{Towards a universal formula for the Bethe 
equations}\label{subsec:universal}

Let us denote in this subsection the affine Lie algebras $\hat g$ in Tables \ref{table: QG symmetry general, 
p=0,...,n} and \ref{table: QG symmetries special cases} by $g^{(t)}$, where $g$ 
is a (non-affine) Lie algebra with rank $\rcal$, and $t=1$ (untwisted) or 
$t=2$ (twisted).\footnote{ The notation $g^{(t)}$ introduced here for 
affine Lie algebras should not be confused with the ``left'' and 
``right'' algebras $g^{(l)}$ and $g^{(r)}$ introduced in Sec. \ref{subsec: Symmetries}.}
The above formulas for the $g^{(t)}$ Bethe equations can be rewritten in a more 
compact form in terms of representation-theoretic quantities 
following \cite{Reshetikhin:1987}:\footnote{For $D_{n+1}^{(2)}$, a 
rescaling $\eta\rightarrow \frac{\eta}{2}$ in (\ref{universal}) is 
necessary in order to match with the Bethe equations as written in 
Sec. \ref{subsec:D2n}.} 
\begin{align}
& \prod_{s=0}^{t-1}\left[\frac{\sinh\left(\frac{u_k^{[l]}}{2}+\left(\lambda_1,\theta^s\alpha_l\right)\eta+\frac{i\pi s}{2}\right)}
{\sinh\left(\frac{u_k^{[l]}}{2}-\left(\lambda_1,\theta^s\alpha_l\right)\eta+\frac{i\pi s}{2}\right)}\right]^{2N}\Phi_{l,p,n}(u_k^{[l]})\nonumber\\
&\quad= \prod_{s=0}^{t-1}\prod_{l^{\prime}=1}^{n}\sideset{}{'}\prod_{j=1}^{m_{l^{\prime}}}
\frac{\sinh\left[\frac{1}{2}\left(u_k^{[l]}-u_j^{[l^{\prime}]}\right)+\left(\alpha_l,\theta^s\alpha_{l^{\prime}}\right)\eta+\frac{i\pi s}{2}\right]}
{\sinh\left[\frac{1}{2}\left(u_k^{[l]}-u_j^{[l^{\prime}]}\right)-\left(\alpha_l,\theta^s\alpha_{l^{\prime}}\right)\eta+\frac{i\pi s}{2}\right]}
\frac{\sinh\left[\frac{1}{2}\left(u_k^{[l]}+u_j^{[l^{\prime}]}\right)+\left(\alpha_l,\theta^s\alpha_{l^{\prime}}\right)\eta+\frac{i\pi s}{2}\right]}
{\sinh\left[\frac{1}{2}\left(u_k^{[l]}+u_j^{[l^{\prime}]}\right)-\left(\alpha_l,\theta^s\alpha_{l^{\prime}}\right)\eta+\frac{i\pi s}{2}\right]} \,, \non\\
&\qquad\qquad\qquad\qquad k = 1, \ldots, m_{l} \,, \qquad l = 1, 
\ldots\,, n \,,
\label{universal}
\end{align}
\noindent
where the product over $j$ has the restriction $(j,l') \ne (k,l)$.
The simple roots $\alpha_i$ of $g$ are given in the orthogonal basis by
\begin{align}
\alpha_i &=e_i-e_{i+1}, \qquad\qquad i=1,...,\rcal-1\,, \nonumber\\
\alpha_{\rcal} &=\begin{cases}
e_{\rcal}-e_{\rcal+1}\hspace{0.9cm} \mbox{for}\quad A_{\rcal}\\
e_{\rcal}\hspace{2.2cm}\mbox{for}\quad B_{\rcal}\\
2e_{\rcal}\hspace{2.02cm}\mbox{for}\quad C_{\rcal}\\
e_{\rcal-1}+e_{\rcal}\hspace{1cm}\mbox{for}\quad D_{\rcal}
\end{cases} \,,
\label{simple roots}
\end{align}
where $e_{i}$ are $\rcal$-dimensional elementary basis vectors (except 
for $A_{\rcal}$, in which case the dimension is $\rcal+1$). The notation $(*\,, *)$ denotes the 
ordinary scalar product, and $\lambda_{1}$ is the first fundamental 
weight of $g$, with $\left( \lambda_{1}\,, \alpha_{i}\right) = 
\delta_{i,1}$. 
For the twisted cases $g^{(2)}$, the order-2 automorphisms $\theta$ of $g$ are given by
\begin{align}
& \theta\alpha_i=\alpha_{2n-i}\,, \quad i = 1, \ldots, 2n-1 
\hspace{0.4cm} \text{for}\quad A_{2n-1}^{(2)}\,, \non \\
& \theta\alpha_i=\alpha_{2n+1-i}\,, \quad i = 1, \ldots, 2n 
\hspace{0.65cm}\text{for}\quad A_{2n}^{(2)}\,, \non \\
& \theta\alpha_i=\alpha_i\,, \quad i = 1, \ldots, n-1\,, \quad  
\theta\alpha_n=\alpha_{n+1} \hspace{0.4cm}\text{for}\quad 
D_{n+1}^{(2)} \,.
\label{automorphism}
\end{align}
The factor $\Phi_{l,p,n}(u)$ in (\ref{universal}) is 
understood to be the appropriate one for $g^{(t)}$, see 
(\ref{phi1}),  (\ref{phi3}),  (\ref{phi2}). It would be interesting 
to also have a universal expression for this factor.

\section{Dynkin labels of the Bethe states}\label{sec:DynkinLabels}

In this section we obtain formulas for the Dynkin labels of the Bethe 
states in terms of the numbers of Bethe roots of each type. Since the 
Dynkin labels of an irrep determine its dimension, these formulas 
help determine the degeneracies of the transfer-matrix eigenvalues.

\subsection{Eigenvalues of the Cartan generators}\label{subsec:Cartan}

We now argue that the eigenvalues of the Cartan generators 
for the Bethe states (\ref{eigenvalueproblem}) are given 
in terms of the cardinalities of the Bethe roots of each type by
\begin{align}
h_i^{(l)} &=m_{p+i-1}-m_{p+i}- \xi\, \delta_{i,n-p} m_{n} -\xi'\, 
    \delta_{i,n-p-1} m_{n}\,, &  i &=1,...,n-p \,, \non \\
h_i^{(r)} &=m_i-m_{i-1}+ \xi\, \delta_{i,n} m_{n} + \xi'\, \delta_{i,n-1} 
m_{n}\,,  & i &=1,...,p \,,
\label{CartanCardinalities}
\end{align}
where $\xi$ and $\xi'$ are given in Table \ref{table:definitions}.

The first step is to compute the asymptotic behavior of $\Lambda^{(m_{1}, 
\ldots, m_{n})}(u,p)$ by computing the expectation value
\be
\langle \Lambda^{(m_{1}, \ldots, m_{n})} | t(u,p) | \Lambda^{(m_{1}, 
\ldots, m_{n})} \rangle  \non
\ee
for $u \rightarrow \infty$. The main idea is to perform a gauge 
transformation to the ``unitary'' gauge \cite{Nepomechie:2018dsn, Nepomechie:2018wzp}, 
so that the asymptotic limit of the monodromy matrices in $t(u,p)$ become 
expressed in terms of the QG generators.  We assume that the Bethe states 
$|\Lambda^{(m_{1}, \ldots\,, m_{n})}\rangle$ are highest-weight states of the 
``left'' algebra
\be
\Delta_{N}(E^{+ (l)}_{i}(p))\, |\Lambda^{(m_{1}, \ldots, m_{n})}\rangle =0 \,, 
\qquad i = 1, \ldots, n-p\,, 
\label{highestweightleft}
\ee
and lowest-weight states of the ``right'' algebra
\be
\Delta_{N}(E^{- (r)}_{i}(p))\, |\Lambda^{(m_{1}, \ldots, m_{n})}\rangle =0 \,, 
\qquad i = 1, \ldots, p\,,
\label{lowesttweightright}
\ee
as is the reference state (\ref{highestweightleftref}), 
(\ref{lowesttweightrightref}). We eventually obtain
\begin{align}
\Lambda^{(m_{1}, \ldots, m_{n})}(u,p) & \sim \sigma(u)\, e^{-2 \kappa\, 
\eta\, N} \Bigg\{ d - 2n + \sum_{j=1}^{p} \left[ \ff^{(r)} e^{4\eta(-j + 
h^{(r)}_{p+1-j})} + \frac{1}{\ff^{(r)}}  e^{-4\eta(-j + 
h^{(r)}_{p+1-j})}\right] \non \\
&\quad +  \sum_{j=p+1}^{n} \left[ \ff^{(l)} e^{-4\eta(n-j + 
h^{(l)}_{j-p})} + \frac{1}{\ff^{(l)}}  e^{4\eta(n-j + 
h^{(l)}_{j-p})}\right] \Bigg\} \quad \text{for} \quad u \rightarrow 
\infty \,,
\label{LambdaAsymCartan}
\end{align}
where
\be
\sigma(u) = \begin{cases}
2^{-2N} e^{2 N u} & \mbox{ for  }  A^{(2)}_{2n-1}\,, A^{(2)}_{2n}\,, 
B^{(1)}_{n}\,, C^{(1)}_{n}\,, D^{(1)}_{n} \\
e^{4 N u} & \mbox{ for  } D^{(2)}_{n+1}
\end{cases} \,,
\label{sigma}
\ee
and
\begin{align}
\ff^{(r)} &=\begin{cases}
   -1     &  \mbox{ for  }  A^{(2)}_{2n}\,, C^{(1)}_{n} \\
e^{4\eta} &  \mbox{ for  }  A^{(2)}_{2n-1}\,, B^{(1)}_{n}\,, D^{(1)}_{n} \\
e^{2\eta} &  \mbox{ for  }  D^{(2)}_{n+1}
\end{cases} \,, \non\\
\ff^{(l)} &=\begin{cases}
   e^{-2\eta}     &  \mbox{ for  }  A^{(2)}_{2n}\,,  B^{(1)}_{n}\,, D^{(2)}_{n+1} \\
-e^{-4\eta} &  \mbox{ for  }  A^{(2)}_{2n-1}\,, C^{(1)}_{n} \\
1 &  \mbox{ for  } D^{(1)}_{n}
\end{cases} \,.
\label{ffs}
\end{align}
Note that the result (\ref{LambdaAsymCartan}) is in terms of the 
eigenvalues of the Cartan generators for the Bethe states.

The second step is to compute again the asymptotic behavior of $\Lambda^{(m_{1}, 
\ldots, m_{n})}(u,p)$, but now using instead the T-Q equation 
(\ref{eigenvalue}). We obtain in this way
\begin{align}
\Lambda^{(m_{1}, \ldots, m_{n})}(u,p) & \sim \sigma(u)\, e^{-2 \kappa\, 
\eta\, N} \Bigg\{ d - 2n + \sum_{l=0}^{n-1} \Big[ \gcal_{l} 
e^{4\eta(l-n + m_{l+1} - m_{l} +\xi \delta_{l,n-1} m_{n} +\xi' 
\delta_{l,n-2} m_{n} )} \non \\
&\quad + \frac{1}{\gcal_{l}}  e^{-4\eta(l-n + m_{l+1} - m_{l} +\xi \delta_{l,n-1} m_{n} +\xi' 
\delta_{l,n-2} m_{n} )}\Big]  \Bigg\} \quad \text{for} \quad u \rightarrow 
\infty \,,
\label{LambdaAsymCardinalities}
\end{align}
where
\be
\gcal_{l}  = \begin{cases}
\ff^{(r)}  e^{4 \eta (n-p)} &  l \le p-1 \\
\ff^{(l)}  e^{4 \eta} &  l \ge p
 \end{cases} \,,
\ee
$\sigma(u)$ is given by (\ref{sigma}), and $\ff^{(r)}, \ff^{(l)}$ are 
given by (\ref{ffs}). Moreover, we define $m_{0}$ as
\be
m_{0} = N \,.
\label{m0}
\ee 
Note that the result (\ref{LambdaAsymCardinalities}) is in terms of the 
cardinalities of the Bethe roots of each type.

Finally, by comparing (\ref{LambdaAsymCartan}) and 
(\ref{LambdaAsymCardinalities}), we obtain the desired result 
(\ref{CartanCardinalities}).

\subsection{Formulas for the Dynkin 
labels}\label{subsec:Dynkin}

The ``left'' Dynkin labels are expressed in terms of the eigenvalues 
of the ``left'' Cartan generators by (see, e.g. \cite{Ahmed:2017mqq, 
Nepomechie:2017hgw})
\begin{align}
a_i^{(l)} &=h_i^{(l)}-h_{i+1}^{(l)}, \qquad\qquad i=1,...,n-p-1\,, \non \\
a_{n-p}^{(l)} &=\begin{cases}
2 h_{n-p}^{(l)}\quad & \mbox{for} \quad g^{(l)}=B_{n-p} \quad \mbox{i.e., 
for}\quad  A^{(2)}_{2n}\,, B^{(1)}_{n}\,, D^{(2)}_{n+1} \\
h_{n-p}^{(l)}\quad & \mbox{for}\quad g^{(l)}=C_{n-p} \quad\mbox{i.e., 
for}\quad  A^{(2)}_{2n-1}\,,   C^{(1)}_{n}\\
h_{n-p-1}^{(l)}+h_{n-p}^{(l)}\quad & \mbox{for}\quad g^{(l)}=D_{n-p} \quad\mbox{i.e., 
for\quad}  D^{(1)}_{n}
\end{cases} \,.
\label{aileft}
\end{align}
Similarly, the ``right'' Dynkin labels are expressed in terms of the eigenvalues 
of the ``right'' Cartan generators by 
\begin{align}
a_i^{(r)} &=-h_i^{(r)}+h_{i+1}^{(r)} \qquad\qquad i=1,...,p-1\,, \non \\
a_p^{(r)} &=\begin{cases}
-2h_p^{(r)}\quad & \mbox{for}\quad g^{(r)}=B_p  \quad\mbox{i.e., 
for}\quad D^{(2)}_{n+1} \\
-h_p^{(r)}\quad & \mbox{for}\quad g^{(r)}=C_p \quad\mbox{i.e., 
for}\quad A^{(2)}_{2n}\,,  C^{(1)}_{n}\\
-h_{p-1}^{(r)}-h_p^{(r)}\quad & \mbox{for}\quad g^{(r)}=D_p \quad\mbox{i.e., 
for}\quad A^{(2)}_{2n-1}\,, B^{(1)}_{n}\,, D^{(1)}_{n}
\end{cases} \,.
\label{airight}
\end{align}
We introduce extra minus signs in (\ref{airight}) (in comparison with 
corresponding formulas in (\ref{aileft}))
since the Bethe 
states are {\em lowest} weights of the ``right'' algebra 
(\ref{lowesttweightright}). The algebras $g^{(l)}$ and 
$g^{(r)}$ for the various affine algebras $\hat g$ are given in 
Table \ref{table: QG symmetry general, p=0,...,n}.

Finally, using the results (\ref{CartanCardinalities}) for the eigenvalues of the Cartan generators 
in terms of the cardinalities of the Bethe roots of each type,
we obtain formulas for the Dynkin labels in terms of the 
cardinalities of the Bethe roots. Explicitly, for the ``left'' Dynkin 
labels ($p=0, 1, \ldots, n-1$):
\begin{align}
a_i^{(l)} &=m_{p+i-1}-2m_{p+i}+m_{p+i+1}\,, 	\label{Dynkinfirst}\\
&\qquad i=1,...,n-p-1 \quad  \mbox{for}\quad A_{2n}^{(2)}\,, 
B_n^{(1)}\,,  D_{n+1}^{(2)}\,, \non \\
&\qquad i=1,...,n-p-2 \quad \mbox{for}\quad  A_{2n-1}^{(2)}\,, 
C_n^{(1)} \,,
\non \\
&\qquad i=1,...,n-p-3 \quad \mbox{for}\quad D_n^{(1)} \,. \non
\end{align}
Moreover, for the values of $i$ not included above:
\begin{align}
& A_{2n}^{(2)}, B_n^{(1)}\,,  D_{n+1}^{(2)}:
	& a_{n-p}^{(l)} &=2m_{n-1}-2m_n \,, \\
& A_{2n-1}^{(2)}, C_n^{(1)}:
	& a_{n-p-1}^{(l)} &=m_{n-2}-2m_{n-1}+2m_n \,, \non \\
	&& a_{n-p}^{(l)} &=m_{n-1}-2m_{n} \,, \\
& D_n^{(1)}:
	& a_{n-p-2}^{(l)} &=m_{n-3}-2m_{n-2}+m_{n-1}+m_n \,, \non \\
	&& a_{n-p-1}^{(l)} &=m_{n-2}-2m_{n-1} \,, \non \\
	&& a_{n-p}^{(l)} &=m_{n-2}-2m_{n} \,.
\end{align}
\noindent
For the ``right'' Dynkin labels  ($p=1, \ldots, n$):
\begin{align}
a_i^{(r)} &=m_{i-1}-2m_i+m_{i+1} \,, \\
&\qquad i=1,...,p-1 \quad \mbox{for} \quad  A_{2n}^{(2)}\,, B_n^{(1)}\,,  D_{n+1}^{(2)} 
\,, \non \\
&\qquad i=1,...,p-2 \quad \mbox{for} \quad  A_{2n-1}^{(2)}\,, 
C_n^{(1)} \non \\
&\qquad i=1,...,p-3 \quad  \mbox{for} \quad  D_n^{(1)}\,. \non
\end{align}
Moreover, for the values of $i$ not included above:
\begin{align}
& A_{2n}^{(2)}:
	& a_p^{(r)} &=m_{p-1}-m_p \,, \\
& B_{n}^{(1)}:
	& a_p^{(r)} &=m_{p-2}-m_p \,, \\
& A_{2n-1}^{(2)}:
	& a_{p-1}^{(r)} &=m_{p-2}-2m_{p-1}+(1+\delta_{p,n})m_p\,, \non \\
	&& a_p^{(r)} &=m_{p-2}-(1+\delta_{p,n}) m_p \,, 
\end{align}
\begin{align}	
& C_{n}^{(1)}:
	& a_{p-1}^{(r)} &=m_{p-2}-2m_{p-1}+(1+\delta_{p,n})m_p \,, 
	\non \\
	&& a_p^{(r)} &=m_{p-1}-(1+\delta_{p,n}) m_p \,, 
\end{align}
\begin{align}	
& D_{n}^{(1)}
	& a_{p-2}^{(r)} &=m_{p-3}-2m_{p-2}+m_{p-1}+\delta_{p,n}m_p 
	\,, \non \\
	&& a_{p-1}^{(r)} 
	&=m_{p-2}-2m_{p-1}+m_p+(\delta_{p,n-1}-\delta_{p,n}) m_n \,, 
	\non \\
	&& a_p^{(r)} &=m_{p-2}-m_p-(\delta_{p,n-1}+\delta_{p,n})m_p 
	\,, \\
& D_{n+1}^{(2)} :
	& a_p^{(r)} &=2m_{p-1}-2m_p \,.
	\label{Dynkinlast}
\end{align}
We remind the reader that $m_{0}$ is defined in (\ref{m0}).

For the cases of overlap with previous results (namely, 
$A_{2n}^{(2)}$ with $p=0,n$ \cite{Ahmed:2017mqq}; $A_{2n-1}^{(2)}$ 
with $p=0,n$ \cite{Nepomechie:2017hgw}; and 
$D_{n+1}^{(2)}$ with $p=n$ \cite{Nepomechie:2017hgw}), the results 
match. 

\subsection{Examples}\label{subsec:ex}

We now illustrate the results of Sec. \ref{subsec:Dynkin} 
 with two simple examples.

\subsubsection{$A_{2n}^{(2)}$ with $n=3$}

As a first example, we consider the case $A_{2n}^{(2)}$ with $n=3$, 
two sites ($N=2$), and either $\varepsilon=0$ or $\varepsilon=1$. 
The four possibilities $p=0, 1, 2, 3$ are 
summarized in Table \ref{example 1}. By solving the Bethe equations 
(see Sec. \ref{subsubsec:BE1})
with a generic value of anisotropy $\eta$, we obtain solutions (not 
shown\footnote{For the cases $p=0$ and $p=n$, such solutions can be 
found in tables in \cite{Ahmed:2017mqq}.}) with 
the values of $m_{1}, m_{2}, m_{3}$ displayed in the table. The 
corresponding Dynkin labels obtained using the formulas from Sec. 
\ref{subsec:Dynkin}, are also displayed in the table. Finally, the 
irreducible representations of the ``left'' and ``right'' algebras
corresponding to these Dynkin labels (obtained e.g. using LieART 
\cite{Feger:2012bs}) are shown in the final column. 
By explicit diagonalization of the transfer matrix, we confirm that the 
degeneracies of the eigenvalues exactly match with the dimensions of the corresponding irreps.

\begin{table}[htb]
    \small
	\centering
	{\renewcommand{\arraystretch}{1}
\begin{tabular}{c|c|c|c|c|c|c|c}
	\hhline{=|=|=|=|=|=|=|=} 
	& $ m_1 $ & $ m_2 $ & $ m_3 $ & $ a_1^{(l)} $ & $ a_2^{(l)} $ & $ a_3^{(l)} $ & Irreps. \\
	\hline
    \multirow{3}{2cm}{\centering{$ p=0 $  $ U_q(B_3) $}}& 0 & 0 & 0 & 2 & 0 & 0 & \textbf{27} \\
    & 1 & 0 & 0 & 0 & 1 & 0 & \textbf{21} \\
    & 2 & 2 & 2 & 0 & 0 & 0 & \textbf{1}  \\
	\hhline{=|=|=|=|=|=|=|=}
	& $ m_1 $ & $ m_2 $ & $ m_3 $ & $ a_1^{(l)} $ & $ a_2^{(l)} $ & $ a_1^{(r)} $ & Irreps.  \\
	\hline
    \multirow{5}{3cm}{\centering{$ p=1 $  $ U_q(B_2)\otimes U_q(C_1) $}}& 0 & 0 & 0 & 0 & 0 & 2 & (\textbf{1,3})\\
	& 1 & 0 & 0 & 1 & 0 & 1 & 2(\textbf{5,2}) \\
	& 2 & 0 & 0 & 2 & 0 & 0 & (\textbf{14,1})\\
	& 2 & 1 & 0 & 0 & 2 & 0 & (\textbf{10,1})\\
    & 2 & 2 & 2 & 0 & 0 & 0 & 2(\textbf{1,1}) \\
	\hhline{=|=|=|=|=|=|=|=}
	& $ m_1 $ & $ m_2 $ & $ m_3 $ & $ a_1^{(l)} $ & $ a_1^{(r)} $ & $ a_2^{(r)} $ & Irreps.  \\
	\hline
	\multirow{6}{3cm}{\centering{$ p=2 $  $ U_q(B_1)\otimes U_q(C_2) $}} & 0 & 0 & 0 & 0 & 2 & 0 & (\textbf{1,10}) \\
	& 1 & 0 & 0 & 0 & 0 & 1 & (\textbf{1,5}) \\
	& 1 & 1 & 0 & 2 & 1 & 0 & 2(\textbf{3,4}) \\
	& 2 & 2 & 0 & 4 & 0 & 0 & (\textbf{5,1}) \\
	& 2 & 2 & 1 & 2 & 0 & 0 & (\textbf{3,1}) \\
	& 2 & 2 & 2 & 0 & 0 & 0 & 2(\textbf{1,1}) \\
	\hhline{=|=|=|=|=|=|=|=} 
	& $ m_1 $ & $ m_2 $ & $ m_3 $ & $ a_1^{(r)} $ & $ a_2^{(r)} $ & $ a_3^{(r)} $ & Irreps.  \\
	\hline
	\multirow{4}{2cm}{\centering{$ p=3 $  $ U_q(C_3) $}}& 0 & 0 & 0 & 2 & 0 & 0 & \textbf{21}\\
	& 1 & 0 & 0 & 0 & 1 & 0 & \textbf{14} \\
	& 1 & 1 & 1 & 1 & 0 & 0 & 2(\textbf{6}) \\
	& 2 & 2 & 2 & 0 & 0 & 0 & 2(\textbf{1}) \\
	\hline
\end{tabular}}
\caption{Numbers of Bethe roots and Dynkin labels for 
$A_{2n}^{(2)}$ with $n=3, N=2$.}
\label{example 1}
\end{table} 

\subsubsection{$D_{n}^{(1)}$ with $n=4$}\label{sec:Dexample}

As a second example, we consider the case $D_{n}^{(1)}$ with $n=4$,  
two sites ($N=2$), and with $\varepsilon=0$. The three cases $p=0, 2, 4$ are 
summarized in Table \ref{example 2}. (We omit the ``special'' cases $p=1, 
3$, whose results are the same as for $p=0, 4$, respectively,
see Table \ref{table: QG symmetries special cases}.)
By solving the Bethe equations 
(see Sec. \ref{subsubsec:BE1})
with a generic value of anisotropy $\eta$, we obtain solutions (not 
shown) with 
the values of $m_{1}, m_{2}, m_{3}, m_{4}$ displayed in the table. The 
corresponding Dynkin labels obtained using the formulas from Sec. 
\ref{subsec:Dynkin}, are also displayed in the table. Finally, the 
irreps of the ``left'' and ``right'' algebras
corresponding to these Dynkin labels are shown in the final column. 

Notice that the values of $m$'s and Dynkin labels for $p=0$ and $p=4$ 
in Table \ref{example 2} are exactly the same, which is due to
the $p \leftrightarrow n-p$ duality (\ref{duality}).

The degeneracy pattern is particularly interesting for the case $p=2$ in Table \ref{example 2}. Indeed, 
by explicitly diagonalizing the transfer matrix for this 
case\footnote{We emphasize that we restrict to generic values of 
$\eta$.}, 
we find the following degeneracies
\begin{equation}
\{1,1,12,16,16,18\} \,.
\end{equation}
 That is, one eigenvalue is repeated 18 times; two 
distinct eigenvalues are each repeated 16 times; etc.
What is happening is that the irreps
\be
(\mathbf{1,9})\,, (\mathbf{9,1})
\label{example 2 terms}
\ee
(see Table \ref{example 2}) are degenerate, thereby giving rise to the 18-fold 
degeneracy, due to the self-duality (\ref{selfdual}). 
Moreover, the irreps
\be
(\mathbf{1,3})\,, (\mathbf{3,1})\,, 
(\mathbf{1,\bar{3}})\,, (\mathbf{\bar{3},1})
\label{example 4 terms}
\ee
(see again Table \ref{example 2}) 
are all degenerate, thereby giving rise to the 12-fold 
degeneracy, due to the self-duality (\ref{selfdual}) and 
$Z_2$ symmetries (\ref{Z2rightprop}), (\ref{Z2leftprop}).

\begin{table}[tb]
    \small
	\centering
	{\renewcommand{\arraystretch}{1.2}
		\begin{tabular}{c|c|c|c|c|c|c|c|c|c}
			\hhline{=|=|=|=|=|=|=|=|=|=} 
			& $ m_1 $ & $ m_2 $ & $ m_3 $ & $ m_4 $ &  $ a_1^{(l)} $ & $ a_2^{(l)} $ & $ a_3^{(l)} $ &  $ a_4^{(l)} $ & Irreps. \\
			\hline
			\multirow{3}{2cm}{\centering{$ p=0 $  $ U_q(D_4) $}} & 0 & 0 & 0 & 0&  2 & 0 & 0 & 0 & \textbf{35} \\
			& 1 & 0 & 0 & 0 & 0 & 1 & 0 & 0 & \textbf{28} \\
			& 2 & 2 & 1 & 1 & 0 & 0 & 0 & 0 & \textbf{1}  \\
			\hhline{=|=|=|=|=|=|=|=|=|=}
			& $ m_1 $ & $ m_2 $ & $ m_3 $ & $ m_4 $ &  $ a_1^{(l)} $ & $ a_2^{(l)} $ & $ a_1^{(r)} $ &  $ a_2^{(r)} $ & Irreps. \\
			\hline
			\multirow{8}{3cm}{\centering{$ p=2 $  $ U_q(D_2)\otimes U_q(D_2) $}}& 0 & 0 & 0 & 0 & 0 & 0 & 2 & 2 & $(\mathbf{1, 9}) $ \rdelim\}{2}{0mm}[18]  \\
			& 2 & 2 & 0 & 0 & 2 & 2 & 0 & 0 & $ (\mathbf{9, 1 }) $ \\
			& 1 & 1 & 0 & 0 & 1 & 1 & 1 & 1 & $ 2(\mathbf{4, 4 }) $ \\
			& 1 & 0 & 0 & 0 & 0 & 0 & 0 & 2 & $ (\mathbf{1, \bar{3} }) $\rdelim\}{4}{0mm}[12] \\
			& 1 & 2 & 1 & 1 & 0 & 0 & 2 & 0 & $ (\mathbf{1, 3 }) $ \\
			& 2 & 2 & 0 & 1 & 2 & 0 & 0 & 0 & $ (\mathbf{3, 1 }) $ \\
			& 2 & 2 & 1 & 0 & 0 & 2 & 0 & 0 & $ (\mathbf{\bar{3}, 1 }) $ \\
			& 2 & 2 & 1 & 1 & 0 & 0 & 0 & 0 & $ 2(\mathbf{1, 1 }) $ \\									
			\hhline{=|=|=|=|=|=|=|=|=|=}
			& $ m_1 $ & $ m_2 $ & $ m_3 $ & $ m_4 $ &  $ a_1^{(r)} $ & $ a_2^{(r)} $ & $ a_3^{(r)} $ &  $ a_4^{(r)} $ & Irreps. \\
			\hline
			\multirow{3}{2cm}{\centering{$ p=4 $  $ U_q(D_4) $}} & 0 & 0 & 0 & 0&  2 & 0 & 0 & 0 & \textbf{35} \\
			& 1 & 0 & 0 & 0 & 0 & 1 & 0 & 0 & \textbf{28} \\
			& 2 & 2 & 1 & 1 & 0 & 0 & 0 & 0 & \textbf{1}  \\
			\hline
	\end{tabular}}
	\caption{Numbers of Bethe roots and Dynkin labels for
	$D_n^{(1)}$ with $n=4, N=2$.}
	\label{example 2}
\end{table} 

For eigenvalues corresponding to more than one irrep, it 
is enough to solve the Bethe equations corresponding to just {\em one} of those 
irreps, such as the irrep with the minimal values of $m$'s. Hence, for the
example (\ref{example 2 terms}), it is enough to consider the 
reference state ($m_{1}=m_{2}=m_{3}=m_{4}=0$). For the example 
(\ref{example 4 terms}), it is enough to consider the 
state with $m_{1}=1, m_{2}=m_{3}=m_{4}=0$. Note that a non-minimal 
set $\{m_{1}, m_{2}, \ldots, m_{n}\}$ generally does {\em not} form a monotonic decreasing sequence, 
i.e. does {\em not}  satisfy $m_{1} \ge m_{2} \ge \ldots \ge 
m_{n}$.\footnote{ It can happen that an eigenvalue 
corresponding to a {\em single} irrep is described by more than one 
set of Bethe roots, and therefore by more than one set of $m$'s; and 
(for some cases with $\frac{n}{2} \le p < n$), the set of
$m$'s corresponding to the Dynkin labels for the irrep may not 
be minimal.
For example, for $C_4^{(1)}$ with $p=3$ and $N=2$, the transfer 
matrix has an eigenvalue with degeneracy 12 and Dynkin labels 
$(a^{(l)}_{1}, a^{(r)}_{1}, a^{(r)}_{2}, a^{(r)}_{3}) = (1, 1, 0, 0)$, which according to the formulas in section \ref{subsec:Dynkin} 
corresponds to $(m_1,m_2,m_3,m_4)=(1,1,1,0)$. Indeed, one can solve
the Bethe equations \eqref{BEgen1}, \eqref{BEgen2}, \eqref{BE extra 
C1}, \eqref{BE extra C2} and find such a solution for this eigenvalue. However, this set of $m$'s
is not minimal, as one can find another solution of these Bethe 
equations for this eigenvalue with only $(m_1,m_2,m_3,m_4)=(1,1,0,0)$.
Another example is $D_4^{(2)}$ with $p=2$ and $N=2$, for which there 
is an eigenvalue with degeneracy 3 and Dynkin labels 
$(a^{(l)}_{1}, a^{(r)}_{1}, a^{(r)}_{2}) = (2, 0, 0)$,
corresponding to $(m_1,m_2,m_3)=(2,2,1)$; but by solving the Bethe equations we
can also find it with $(m_1,m_2,m_3)=(2,1,1)$.}

For $\varepsilon=1$ (and still $n=4, p=2$), the transfer matrix has an
additional ``bonus'' symmetry \cite{Nepomechie:2018dsn}.
Consequently, the two irreps $(\mathbf{4,4})$ in Table \ref{example 2}
become degenerate (giving rise to a 32-fold degeneracy), and the two
irreps $(\mathbf{1,1})$ become degenerate (giving rise to a 2-fold
degeneracy).  Interestingly,
these levels have the
singular (exceptional) Bethe roots $u^{(1)}=2\eta\,, 
u^{(2)}=4\eta$; and for the 2-fold degenerate level, these Bethe roots 
are repeated. 
This phenomenon is discussed further in 
Appendix \ref{sec:bonus}.

\section{Duality and the Bethe ansatz}\label{sec:dualityBA}

For the cases $C^{(1)}_{n}$, $D^{(1)}_{n}$ and $D^{(2)}_{n+1}$, the
$p \leftrightarrow n-p$ duality property of the transfer matrix (\ref{duality}) is reflected 
in the Bethe ansatz solution. For concreteness, we restrict our 
attention here to the case $C^{(1)}_{n}$, for which 
\be
f(u,p) = -\phi(u,p) \,,
\label{ffunc}
\ee
where $\phi(u,p)$ is given by (\ref{phi}).

The duality property of the transfer matrix (\ref{duality}) implies 
that corresponding eigenvalues satisfy
\be
\Lambda(u,p) = f(u,p)\, 
\Lambda(u,n-p)  \,.
\label{dualityeig}
\ee 
Let us define the rescaled eigenvalue $\lambda(u,p)$ such that
\be
\Lambda(u,p) = \phi(u,p)\, 
\lambda(u,p) \,.
\label{barLambdadef}
\ee
In terms of $\lambda(u,p)$, the duality relation (\ref{dualityeig}) 
takes the form 
\be
\lambda(u,p) = \frac{1}{f(u,p)}\, 
\lambda(u,n-p)  \,,
\label{dualitybarLambda}
\ee 
as follows from (\ref{ffunc}), (\ref{barLambdadef}) and $f(u,n-p) = 
1/f(u,p)$.

Let us now try to understand how the duality relation 
(\ref{dualitybarLambda}) emerges from the Bethe 
ansatz solution (\ref{eigenvalue}), which in terms of $\lambda(u,p)$ (\ref{barLambdadef}) 
reads
\begin{align}
\lambda(u,p)=&A(u)\, z_0(u)\, y_0(u,p)\, c(u)^{2N}+\tilde{A}(u)\, \tilde{z}_0(u)\, \tilde{y}_0(u,p)\, \tilde{c}(u)^{2N}\nonumber\\
&\hspace{-0.3in} +\Big\{\sum_{l=1}^{n-1}\left[z_l(u)\, y_l(u,p)\, B_l(u)+\tilde{z}_l(u)\, 
\tilde{y}_l(u,p)\, \tilde{B}_l(u)\right] \Big\}\, b(u)^{2N}\,.
\label{barLambda}
\end{align}
For the self-dual case $p=n/2$, the relation (\ref{dualitybarLambda}) is 
obvious, since $f(u,n/2) = 1$. For the case $p=0$, we note the 
identity
\be
\frac{y_{l}(u, 0)}{y_{l}(u, n)} = \frac{1}{f(u,0)} \,, \qquad l = 0, 1, 
\ldots, n-1\,.
\label{p0id}
\ee
Since $A(u)$ and $\{ B_{l}(u) \}$ for $p=0$ are the same as for $p=n$ 
(the Bethe equations for $p=0$ are the same as for $p=n$),
it follows from (\ref{barLambda}) and (\ref{p0id}) that
\be
\lambda(u,0) = \frac{1}{f(u,0)}\, 
\lambda(u,n)  \,,
\ee 
in agreement with (\ref{dualitybarLambda}).

To derive the duality relation (\ref{dualitybarLambda}) from the Bethe ansatz solution 
for $0 < p < n/2$ requires more effort. For simplicity, let us 
consider as an example the case $n=3$ with $p=1$, which is related by 
duality to $p=2$. The rescaled eigenvalue is given by (\ref{barLambda})
\begin{align}
\lambda(u,p) & = z_0(u)\, y_0(u,p)\, \frac{Q^{[1]}(u+2\eta)}{Q^{[1]}(u-2\eta)}\, 
\left[2 \sinh(\frac{u}{2}-2\eta) 
\sinh(\frac{u}{2}-8\eta)\right]^{2N} \nonumber\\
& + \Big\{ z_1(u)\, 
y_1(u,p)\,\frac{Q^{[1]}(u-6\eta)}{Q^{[1]}(u-2\eta)}\frac{Q^{[2]}(u)}{Q^{[2]}(u-4\eta)}   \nonumber\\
&+ z_2(u)\, y_2(u,p)\, \frac{Q^{[2]}(u-8\eta)}{Q^{[2]}(u-4\eta)} 
\frac{Q^{[3]}(u)}{Q^{[3]}(u-8\eta)} 
\Big\}\, \left[2 \sinh(\frac{u}{2}) 
\sinh(\frac{u}{2}-8\eta)\right]^{2N} + \ldots \,,
\label{barLambdan3}
\end{align}
where the crossed terms (indicated by the ellipsis) have not been 
explicitly written. Let us define the barred Q-functions  

\begin{equation}
\bar{Q}^{[l]}(u)=\prod_{j=1}^{\bar{m}_l}\sinh\left(\tfrac{1}{2}(u-\bar{u}_j^{[l]})\right)\sinh\left(\tfrac{1}{2}(u+\bar{u}_j^{[l]})\right)\,,
\qquad  \bar{Q}^{[l]}(-u) = \bar{Q}^{[l]}(u) \,,
\label{barQ}
\end{equation}
(in terms of unbarred ones $Q^{[l]}(u)$) as follows:
\begin{align}
S(u) - S(-u) &= \ccal\,  \sinh^{2N}(\frac{u}{2})\, \sinh(u)\, 
Q^{[2]}(u) \,, \label{d1}  \\
S(u) &= \chi(u+2\eta)\, Q^{[1]}(u+2\eta)\, \bar{Q}^{[1]}(u-2\eta) \,, 
\label{d2}  \\
\bar{Q}^{[2]}(u)  &= Q^{[2]}(u)  \,, \label{d3}  \\
\bar{Q}^{[3]}(u)  &= Q^{[3]}(u)  \,, \label{d4}
\end{align}
where 
\be
\chi(u)  = 1 + \cosh(u) \,, \qquad \ccal = 2 
\sinh(2\eta(1+2m_{1}-m_{2}-N)) \,,
\ee
and 
\be
    \bar{m}_{1} = N - m_{1} + m_{2} \,,  \qquad 
    \bar{m}_{2} = m_{2} \,, \qquad
    \bar{m}_{3} = m_{3} \,.
    \label{mbarmrltn}
\ee
(The above results for $\bar{m}_{1}$ and $\ccal$ follow from the 
asymptotic limit $u\rightarrow \infty$ of (\ref{d1}).) We show below that, if $Q^{[l]}(u)$ are the 
Q-functions for $p=1$, then $\bar{Q}^{[l]}(u)$ are the Q-functions 
for $p=2$. 

\subsection{Duality of the Bethe equations}

We first show that (\ref{d1})-(\ref{d4}) map the $p=1$ Bethe equations:
\begin{align}
\left[\frac{\sinh\left(\frac{u_k^{[1]}}{2}+\eta\right)}
           {\sinh\left(\frac{u_k^{[1]}}{2}-\eta\right)}\right]^{2N}
\left[\frac{\cosh\left(\frac{u_k^{[1]}}{2}-2\eta\right)}
           {\cosh\left(\frac{u_k^{[1]}}{2}+2\eta\right)}\right]^{2} &=
\frac{Q_k^{[1]}\left(u_k^{[1]}+4\eta\right)}{Q_k^{[1]}\left(u_k^{[1]}-4\eta\right)}\frac{Q^{[2]}\left(u_k^{[1]}-2\eta\right)}{Q^{[2]}
\left(u_k^{[1]}+2\eta\right)}\,, \label{BEC3p11}\\
1 &=\frac{Q^{[1]}\left(u_k^{[2]}-2\eta\right)}{Q^{[1]}\left(u_k^{[2]}+2\eta\right)}
\frac{Q_k^{[2]}\left(u_k^{[2]}+4\eta\right)}{Q_k^{[2]}\left(u_k^{[2]}-4\eta\right)}
\frac{Q^{[3]}\left(u_k^{[2]}-2\eta\right)}{Q^{[3]}\left(u_k^{[2]}+2\eta\right)}\,, \label{BEC3p12}\\
1 &=\frac{Q^{[2]}\left(u_k^{[3]}-4\eta\right)}{Q^{[2]}\left(u_k^{[3]}+4\eta\right)}
\frac{Q_k^{[3]}\left(u_k^{[3]}+8\eta\right)}{Q_k^{[3]}\left(u_k^{[3]}-8\eta\right)} \,, \label{BEC3p13}
\end{align} 
to the $p=2$ Bethe equations:
\begin{align}
\left[\frac{\sinh\left(\frac{\bar{u}_k^{[1]}}{2}+\eta\right)}
           {\sinh\left(\frac{\bar{u}_k^{[1]}}{2}-\eta\right)}\right]^{2N}
 &=
\frac{\bar{Q}_k^{[1]}\left(\bar{u}_k^{[1]}+4\eta\right)}{\bar{Q}_k^{[1]}\left(\bar{u}_k^{[1]}-4\eta\right)}\frac{\bar{Q}^{[2]}\left(\bar{u}_k^{[1]}-2\eta\right)}{\bar{Q}^{[2]}
\left(\bar{u}_k^{[1]}+2\eta\right)}\,, \label{BEC3p21}\\
\left[\frac{\cosh\left(\frac{\bar{u}_k^{[2]}}{2}-\eta\right)}
           {\cosh\left(\frac{\bar{u}_k^{[2]}}{2}+\eta\right)}\right]^{2} &=\frac{\bar{Q}^{[1]}\left(\bar{u}_k^{[2]}-2\eta\right)}{\bar{Q}^{[1]}\left(\bar{u}_k^{[2]}+2\eta\right)}
\frac{\bar{Q}_k^{[2]}\left(\bar{u}_k^{[2]}+4\eta\right)}{\bar{Q}_k^{[2]}\left(\bar{u}_k^{[2]}-4\eta\right)}
\frac{\bar{Q}^{[3]}\left(\bar{u}_k^{[2]}-2\eta\right)}{\bar{Q}^{[3]}\left(\bar{u}_k^{[2]}+2\eta\right)}\,, \label{BEC3p22}\\
1 &=\frac{\bar{Q}^{[2]}\left(\bar{u}_k^{[3]}-4\eta\right)}{\bar{Q}^{[2]}\left(\bar{u}_k^{[3]}+4\eta\right)} 
\frac{\bar{Q}_k^{[3]}\left(\bar{u}_k^{[3]}+8\eta\right)}{\bar{Q}_k^{[3]}\left(\bar{u}_k^{[3]}-8\eta\right)} \,, \label{BEC3p23}
\end{align} 
Evidently, it follows from (\ref{d3}) and (\ref{d4}) that 
(\ref{BEC3p13}) implies (\ref{BEC3p23}). 

Setting $u=u_{k}^{[2]}$ in 
(\ref{d1}), remembering that $Q^{[2]}(u_{k}^{[2]})=0$, we obtain
the relation
\be
\frac{Q^{[1]}\left(u_k^{[2]}-2\eta\right)}{Q^{[1]}\left(u_k^{[2]}+2\eta\right)} = 
\frac{\chi\left(u_k^{[2]}+2\eta\right)}{\chi\left(u_k^{[2]}-2\eta\right)}
\frac{\bar{Q}^{[1]}\left(u_k^{[2]}-2\eta\right)}{\bar{Q}^{[1]}\left(u_k^{[2]}+2\eta\right)} \,.
\ee
With the help of this relation, it follows that (\ref{BEC3p12}) 
implies (\ref{BEC3p22}).

Setting $u= \pm u_{k}^{[1]} + 2\eta$ in (\ref{d1}), noting
that therefore $Q^{[1]}(u- 2\eta)=0$ and $S(-u)=0$, we obtain the pair of 
relations
\begin{align}
\chi(u_k^{[1]}+4\eta)\, Q^{[1]}(u_k^{[1]}+4\eta)\, \bar{Q}^{[1]}(u_k^{[1]})
& = \ccal\, \sinh^{2N}(\frac{u_k^{[1]}}{2}+\eta)\, 
\sinh(u_k^{[1]}+2\eta)\, 
Q^{[2]}(u_k^{[1]}+2\eta) \,, \non \\
\chi(u_k^{[1]}-4\eta)\, Q^{[1]}(u_k^{[1]}-4\eta)\, \bar{Q}^{[1]}(u_k^{[1]})
& = -\ccal\, \sinh^{2N}(\frac{u_k^{[1]}}{2}-\eta)\, 
\sinh(u_k^{[1]}-2\eta)\, 
Q^{[2]}(u_k^{[1]}-2\eta) \,.
\end{align}
Forming the ratio of these relations, we arrive at the Bethe equation (\ref{BEC3p11}). 
Similarly, setting  $u= \pm \bar{u}_{k}^{[1]} - 2\eta$ in (\ref{d1}), 
we obtain the Bethe equation (\ref{BEC3p21}).

\subsection{Duality of the transfer-matrix 
	eigenvalues}

In order to relate the transfer-matrix eigenvalues for $p=1$ and $p=2$, we observe 
from (\ref{d1}) that
\be
\ccal = \frac{S(u) - S(-u)}{\sinh^{2N}(\frac{u}{2})\, \sinh(u)\, Q^{[2]}(u)}
= \frac{S(u-4\eta) - S(-u+4\eta)}{\sinh^{2N}(\frac{u}{2}-2\eta)\, 
\sinh(u-4\eta)\, Q^{[2]}(u-4\eta)}
\,, 
\ee 
where the second equality follows from shifting $u \mapsto u 
-4\eta$. Making use of (\ref{d2}) and (\ref{d3}), and rearranging 
terms, we obtain the relation
\begin{align}
\MoveEqLeft \sinh^{2N}(\frac{u}{2}-2\eta)\, \sinh(u-4\eta)\, \chi(u+2\eta)\, 
\frac{Q^{[1]}(u+2\eta)}{Q^{[1]}(u-2\eta)} \non\\
& + \sinh^{2N}(\frac{u}{2})\, \sinh(u)\, \chi(u-6\eta)\, 
\frac{Q^{[1]}(u-6\eta)}{Q^{[1]}(u-2\eta)}\frac{Q^{[2]}(u)}{Q^{[2]}(u-4\eta)}\non \\
&  = 
\sinh^{2N}(\frac{u}{2}-2\eta)\, \sinh(u-4\eta)\, \chi(u-2\eta)\, 
\frac{\bar{Q}^{[1]}(u+2\eta)}{\bar{Q}^{[1]}(u-2\eta)}  \non\\
& + \sinh^{2N}(\frac{u}{2})\, \sinh(u)\, \chi(u-2\eta)\, 
\frac{\bar{Q}^{[1]}(u-6\eta)}{\bar{Q}^{[1]}(u-2\eta)}\frac{\bar{Q}^{[2]}(u)}{\bar{Q}^{[2]}(u-4\eta)}  \,.
\end{align}
This relation implies that
\begin{align}
\MoveEqLeft z_0(u)\, y_0(u,1)\, \frac{Q^{[1]}(u+2\eta)}{Q^{[1]}(u-2\eta)}\, \left[2 \sinh(\frac{u}{2}-2\eta) 
\sinh(\frac{u}{2}-8\eta)\right]^{2N} \non\\
& + z_1(u)\, y_1(u,1)\,\frac{Q^{[1]}(u-6\eta)}{Q^{[1]}(u-2\eta)}\frac{Q^{[2]}(u)}{Q^{[2]}(u-4\eta)} \left[2 \sinh(\frac{u}{2}) 
\sinh(\frac{u}{2}-8\eta)\right]^{2N} \non\\
& =  \frac{1}{f(u,1)} \Big\{ z_0(u)\, y_0(u,2)\, \frac{\bar{Q}^{[1]}(u+2\eta)}{\bar{Q}^{[1]}(u-2\eta)} \, \left[2 \sinh(\frac{u}{2}-2\eta) 
\sinh(\frac{u}{2}-8\eta)\right]^{2N} \non\\
& + z_1(u)\, y_1(u,2)\,\frac{\bar{Q}^{[1]}(u-6\eta)}{\bar{Q}^{[1]}(u-2\eta)}\frac{\bar{Q}^{[2]}(u)}{\bar{Q}^{[2]}(u-4\eta)} \left[2 \sinh(\frac{u}{2}) 
\sinh(\frac{u}{2}-8\eta)\right]^{2N} \Big\}
\,.
\end{align}
Finally, in view of also (\ref{barLambdan3}), (\ref{d3}), (\ref{d4}) 
and the identity
\be
\frac{y_{2}(u, 1)}{y_{2}(u, 2)} = \frac{1}{f(u,1)} \,,
\ee
we conclude that the duality relation (\ref{dualitybarLambda}) is indeed 
satisfied by the Bethe ansatz solution for $n=3\,, p=1$.

\subsection{Duality of the Dynkin labels}

It is interesting to see if the formulas in Sec.  \ref{subsec:Dynkin}
for the Dynkin labels are compatible with duality.  For the case $n=3,
p=1$, where the QG symmetry is $U_{q}(C_{2}) \otimes U_{q}(C_{1})$, 
the Dynkin labels are given by
\begin{align}
a^{(l)}_{1} &= m_{1} - 2m_{2} + 2m_{3} \,, \non \\
a^{(l)}_{2} &= m_{2} -  2m_{3} \,, \non \\
a^{(r)}_{1} &= N -  m_{1} \,.
\label{Dynkinp1}
\end{align}
On the other hand, for the dual case $n=3,
p=2$, where the QG symmetry is $U_{q}(C_{1}) \otimes U_{q}(C_{2})$, 
the Dynkin labels are given by
\begin{align}
\bar{a}^{(l)}_{1} &= \bar{m}_{2} - 2\bar{m}_{3} \,, \non \\
\bar{a}^{(r)}_{1} &= N - 2\bar{m}_{1} + \bar{m}_{2} \,, \non \\
\bar{a}^{(r)}_{2} &= \bar{m}_{1} - \bar{m}_{2} \,,
\label{Dynkinp2}
\end{align}
where we again use a bar to denote quantities for the $p=2$ case. If a 
transfer-matrix eigenvalue ($\Lambda(u,1)$ or equivalently its dual $\Lambda(u,2)$)
forms a single irreducible representation of the QG, then we 
expect that the corresponding Dynkin labels (\ref{Dynkinp1}) and (\ref{Dynkinp2}) 
should be related by the duality relations\footnote{For general values of $n$ and $p$, we 
expect the duality relations
\begin{align*}
\bar{a}^{(l)}_{i} &= a^{(r)}_{i} \,, \qquad i = 1, \ldots, p\,,  \\
\bar{a}^{(r)}_{i} &= a^{(l)}_{i} \,, \qquad i = 1, \ldots, n-p \,,
\end{align*}
where the unbarred and barred quantities correspond to $p$ and $n-p$, 
respectively.}
\begin{align}
\bar{a}^{(l)}_{1} &= a^{(r)}_{1} \,, \non \\
\bar{a}^{(r)}_{i} &= a^{(l)}_{i} \,, \qquad i = 1, 2 \,.
\label{Dynkinrelation}
\end{align}
Making use of the relation (\ref{mbarmrltn}) between $\{ m_{l} \}$ and $\{ 
\bar{m}_{l} \}$, we find that the 
relations (\ref{Dynkinrelation}) are indeed satisfied, provided that 
the $m$'s satisfy the constraint
\be
N = m_{1} + m_{2} - 2m_{3}  \qquad  \text{or equivalently} \qquad \bar{m}_{1} - 2\bar{m}_{2} + 2 
\bar{m}_{3} = 0
\,.
\label{contraint}
\ee
Some simple examples for $N=2$ are displayed in Table \ref{table:Dynkindual}.

\begin{table}[htb]
	\centering
	{\renewcommand{\arraystretch}{1.2}
		\begin{tabular}{c|c|c|c|c|c|c|c}
			\hhline{=|=|=|=|=|=|=|=} 
			& $m_1$ & $m_2$ & $m_3$ & $a_1^{(l)}$ & $a_2^{(l)}$  &  $a_1^{(r)}$ & Irreps. \\
			\hline
			\multirow{2}{*}{\centering{$p=1$   
			$U_q(C_2) \otimes U_q(C_1)$}} & 2 & 0 & 0 &  
			2 & 0 & 0 & $(\mathbf{10, 1})$ \\
			 & 2 & 2 & 1 &  0 & 0 & 0 & $2(\mathbf{1, 1})$ \\
			\hhline{=|=|=|=|=|=|=|=|}
			& $\bar{m}_1$ & $\bar{m}_2$ & $\bar{m}_3$ &  
			$\bar{a}_1^{(l)}$ & 
			$\bar{a}_1^{(r)}$  &  $\bar{a}_2^{(r)}$ & Irreps. \\
			\hline
			\multirow{2}{*}{\centering{$p=2$   
			$U_q(C_1) \otimes U_q(C_2)$}} & 0 & 0 & 0 &  
			0 & 2 & 0 &  $(\mathbf{1, 10})$ \\
			& 2 & 2 & 1 & 0 & 0 & 0 &  $2(\mathbf{1, 1})$ \\
			\hline
	\end{tabular}}
	\caption{Numbers of Bethe roots, which satisfy the 
	constraint (\ref{contraint}), and the corresponding Dynkin labels for
	$C_n^{(1)}$ with $n=3, N=2$ and $p=1, 2$.}
	\label{table:Dynkindual}
\end{table} 

Interestingly, not all transfer-matrix eigenvalues have Bethe roots that satisfy the constraint 
(\ref{contraint}). (A simple example is the reference-state eigenvalue, for 
which $m_{1}=m_{2}=m_{3}=0$.)
Such transfer-matrix eigenvalues correspond to {\em reducible}
representations of the QG (i.e., they correspond to a direct sum of 
two or more irreps). Indeed, it was noted in 
\cite{Nepomechie:2018dsn} (see Sec. 6.4.2) that 
for $C^{(1)}_{n}$ with odd $n$ and $p=\frac{n \pm 1}{2}$, there are 
additional degeneracies in the spectrum, which may be due to some yet 
unknown discrete symmetry.

\subsection{Further remarks}

We have seen that, for the case $C^{(1)}_{n}$ with $n=3$, the 
relations (\ref{d1})-(\ref{d4}) implement the duality transformation 
$p=1 \leftrightarrow p=2$ on the Bethe ansatz solution. Note that the 
Bethe roots corresponding to transfer-matrix eigenvalues related by 
this duality satisfy $u_{k}^{[2]} = \bar{u}_{k}^{[2]}$ and $u_{k}^{[3]} = 
\bar{u}_{k}^{[3]}$; i.e. only the type-1 Bethe roots ($u_{k}^{[1]}, 
\bar{u}_{k}^{[1]}$) are different. We expect that, for $C^{(1)}_{n}$ 
with other values of $n$, as well as for $D^{(1)}_{n}$ and 
$D^{(2)}_{n+1}$, generalizations of the relations (\ref{d1})-(\ref{d4}) can be 
found to implement the duality transformations 
$p \leftrightarrow n-p$ on the Bethe ansatz solutions. For
supersymmetric (graded) integrable spin chains, a different type
of ``duality'' transformation can be defined,  which can be 
implemented on the corresponding Bethe ansatz solutions by relations 
somewhat analogous to (\ref{d1})-(\ref{d4}), see e.g.  
\cite{Gohmann:2003, Beisert:2005di} and references therein.

\section{Discussion}\label{sec:discussion}

We have proposed Bethe ansatz solutions for several infinite families
of integrable open quantum spin chains with QG symmetry that were
identified in \cite{Nepomechie:2018dsn, Nepomechie:2018wzp}.  
In particular, we have found that the Bethe equations take the simple 
form (\ref{universal}), where the factor $\Phi_{l,p,n}(u)$, which is 
different from 1 only if $l=p$, is given by (\ref{phi1}),  (\ref{phi3}),  (\ref{phi2}). 
We have also
proposed formulas for the Dynkin labels of the Bethe states in terms of
the numbers of Bethe roots of each type, see Eqs. (\ref{Dynkinfirst}) 
- (\ref{Dynkinlast}). Finally, we have initiated an 
investigation of how the duality transformations (\ref{duality}) are 
implemented on the Bethe ansatz 
solutions, see (\ref{d1})-(\ref{d4}).

We mention here a few of the many interesting problems that remain to be
addressed.  It would be desirable to use nested algebraic Bethe ansatz
(see e.g. \cite{Li:2005pp, Li:2006mv}) to rederive the Bethe ansatz
solutions,  to obtain the Bethe states,
and to prove the highest/lowest weight conjectures
(\ref{highestweightleft}), (\ref{lowesttweightright}).  However, the
latter computation would require using the reference state
(\ref{reference}), which would in turn require a set of creation
operators different from those used in \cite{Li:2005pp, Li:2006mv}.

It would be interesting to find a Bethe ansatz solution 
for the case $D^{(2)}_{n+1}$ with $\varepsilon=1$ (we considered in 
Sec. \ref{subsec:D2n} only $\varepsilon=0$), to find a 
completely universal form of the Bethe equations for the QG-invariant models considered 
here (see Sec. \ref{subsec:universal}), to further 
investigate how Bethe ansatz solutions transform under duality (we 
focused in Sec. \ref{sec:dualityBA} primarily on the case 
$C^{(1)}_{3}$),
and to understand the 
connection between bonus symmetry and singular solutions of the Bethe 
equations (see Appendix \ref{sec:bonus}). It would also be interesting to investigate the rational limit 
of these models, and to compare with results in the literature e.g. 
\cite{Gombor:2017qsy}.

\section*{Acknowledgments}
RN was supported in part by a Cooper fellowship.  He gratefully
acknowledges support from the Simons Center for Geometry and Physics,
Stony Brook University, where some of this work was performed
during the Exactly Solvable Program.
ALR was supported by the S\~ao Paulo
Research Foundation FAPESP under the process  \# 2017/03072-3 and \# 
2015/00025-9. ALR thanks the University of Miami for its warm hospitality.

\appendix

\section{Bonus symmetry and singular solutions}\label{sec:bonus}

For the cases $C_n^{(1)}\,, D_n^{(1)}\,, D_{n+1}^{(2)}$ with 
$p=\frac{n}{2}$ ($n$ even) and $\varepsilon=1$, the transfer matrix has a 
``bonus'' symmetry (i.e., a symmetry in addition to self-duality), 
leading to higher degeneracies in comparison with $\varepsilon=0$ \cite{Nepomechie:2018dsn, 
Nepomechie:2018wzp}. We observe here that the solutions of the Bethe equations corresponding  
to such degenerate levels are singular (exceptional).

As an example, we consider the case $C_n^{(1)}$ with $n=2, p=1$. 
From the $U_{q}(C_{1}) \otimes U_{q}(C_{1})$ symmetry of the transfer 
matrix, we expect (for generic values of $\eta$) the following 
Hilbert space decompositions
\begin{align}
N=2:\quad &\left[(\mathbf{2,1})\oplus (\mathbf{1,2})\right]^{\otimes 2}=2(\mathbf{1,1})\oplus 
2(\mathbf{2,2})\oplus (\mathbf{3,1})\oplus (\mathbf{1,3})\,, \label{decomp2}\\
N=3:\quad &\left[(\mathbf{2,1})\oplus (\mathbf{1,2})\right]^{\otimes 3}=5(\mathbf{2,1})\oplus 
5(\mathbf{1,2})\oplus 3(\mathbf{3,2})\oplus 3(\mathbf{2,3})\oplus (\mathbf{4,1})\oplus (\mathbf{1,4}) \,.
\label{decomp3}
\end{align}
\noindent
However, by diagonalizing the transfer matrix directly, we observe the 
following degeneracy patterns
\begin{align}
N=2:\quad&  \{1,1,4,4,6\}\hspace{2.58cm}\text{when 
}\varepsilon=0\,,\label{deg2e0}\\
&\{2,8,6\}\hspace{3.35cm}\text{when }\varepsilon=1\,, \label{deg2e1}\\
N=3:\quad & \{4,4,4,4,4,8,12,12,12\}\quad \text{when }\varepsilon=0 
\,,\label{deg3e0}\\
&\{4,8,8,8,12,24\}\hspace{1.8cm}\text{when }\varepsilon=1 
\,.\label{deg3e1}
\end{align}

Let us first consider the case $N=2$.
Comparing the decomposition \eqref{decomp2} with the degeneracies for 
$\varepsilon=0$ \eqref{deg2e0}, we see that they do not completely match: the 
$(\mathbf{3,1})$ and $(\mathbf{1,3})$ are degenerate (thereby giving rise to the 6-fold 
degeneracy) due to the self-duality (\ref{selfdual}). However, the 
degeneracies for $\varepsilon=1$ \eqref{deg2e1} are even higher: the 
two $(\mathbf{2,2})$ are degenerate (thereby giving rise to the 8-fold 
degeneracy) and the two $(\mathbf{1,1})$ are degenerate (thereby giving rise 
to the 2-fold degeneracy) due to the ``bonus'' symmetry.

The key new point is that, among the Bethe roots 
corresponding to the levels with 8-fold degeneracy and 2-fold 
degeneracy,
is the exact Bethe root $u^{[1]}=2\eta$ (which is repeated for the 
2-fold degenerate level), for which the Bethe equations 
have a zero or pole.

The bonus symmetry is also present for $N=3$, see \eqref{decomp3},  
\eqref{deg3e0}, \eqref{deg3e1}. The levels that are degenerate due to the 
bonus symmetry (namely, the level with 24-fold degeneracy, and two 
levels with 8-fold degeneracy) again contain the singular 
solution $u^{[1]}=2\eta$, which is repeated for the 8-fold 
degenerate levels.

For all the examples that we have checked (another example is noted 
in Sec. \ref{sec:Dexample}), singular solutions occur
if and only if the states are affected by the bonus symmetry. 
However, a general understanding of this phenomenon is still lacking.

\section{Bethe ansatz solutions for some additional cases}\label{sec:extra}
 
In the main part of this paper, we do {\em not} consider
the K-matrices \eqref{KRa} for
the cases $A_{2n-1}^{(2)}$ and $B_n^{(1)}$ with $p=1$, and 
$D_n^{(1)}$ with $p= 1\,, n-1$, as emphasized in \eqref{special cases}. These K-matrices are excluded 
because the corresponding transfer matrices do {\em not} have QG 
symmetry corresponding to removing one node from the Dynkin diagram.
(This is the reason why we consider instead the K-matrices 
(\ref{Kspecialp1}) and (\ref{Kspecialpn1}) for these cases.)
Nevertheless, the transfer matrices for these cases are integrable, 
and we have also determined their spectra.  We briefly note here the Bethe
ansatz solutions for these cases.
 
For these cases (i.e., for the transfer matrices constructed using the 
K-matrices \eqref{KRa} for $A_{2n-1}^{(2)}$ and $B_n^{(1)}$ with $p=1$, and 
for $D_n^{(1)}$ with $p= 1\,, n-1$), the transfer matrix eigenvalues 
are in fact given by (\ref{eigenvalue}), where the functions
$y_l(u,p)$ are given by \eqref{yl}, \eqref{G1}, \eqref{F1}.
Hence, the Bethe equations for 
$A_{2n-1}^{(2)}$, $B_n^{(1)}$ and $D_n^{(1)}$ with $p=1$ are again 
those in Sec. \ref{subsubsec:BE1}, with the functions $\Phi_{l,p,n}$ 
given by \eqref{phi1}. 

For  $D_n^{(1)}\, (n>3)$  with $p=n-1$, the
Bethe equations for $l\leq n-2$ are the ones given in
\eqref{BEgen1},\eqref{BEgen2},\eqref{BE extra D1}; but 
the Bethe equations for $l=n-1, n$ are given by 
\be
\left[\frac{\cosh\left(\frac{u_k^{[n-1]}}{2}+\eta+\frac{i \pi
	\varepsilon}{2}\right)}{\cosh\left(\frac{u_k^{[n-1]}}{2}-\eta+\frac{i \pi
	\varepsilon}{2}\right)}\right]^2 =\frac{Q^{[n-2]}\left(u_k^{[n-1]}-2\eta\right)}
	{Q^{[n-2]}\left(u_k^{[n-1]}+2\eta\right)}\frac{Q_k^{[n-1]}\left(u_k^{[n-1]}+4\eta\right)}{Q_k^{[n-1]}\left(u_k^{[n-1]}-4\eta\right)}\,,
\ee
\be
\left[\frac{\cosh\left(\frac{u_k^{[n]}}{2}+\eta+\frac{i \pi
		\varepsilon}{2}\right)}{\cosh\left(\frac{u_k^{[n]}}{2}-\eta+\frac{i \pi
		\varepsilon}{2}\right)}\right]^2 =\frac{Q^{[n-2]}\left(u_k^{[n]}-2\eta\right)}
		{Q^{[n-2]}\left(u_k^{[n]}+2\eta\right)}\frac{Q_k^{[n]}\left(u_k^{[n]}+4\eta\right)}{Q_k^{[n]}\left(u_k^{[n]}-4\eta\right)} \,,
\label{nonQG}
\ee
instead of by \eqref{BE extra D2} and \eqref{BE extra D3}. 
 In contrast with the QG-invariant case, the LHS of \eqref{nonQG}
has a nontrivial ($\ne 1$) factor, even though $l=n\ne p$.


\providecommand{\href}[2]{#2}\begingroup\raggedright\endgroup

\end{document}